\documentstyle[12pt]{article}
\newcommand{\nc}{\newcommand}

\nc{\dbar}{\bar{\partial}}

\catcode`\@=11

\@addtoreset{equation}{section}
\def\theequation{\thesection\arabic{equation}}

\def\@normalsize{\@setsize\normalsize{15pt}\xiipt\@xiipt
\abovedisplayskip 14pt plus3pt minus3pt%
\belowdisplayskip \abovedisplayskip
\abovedisplayshortskip  \z@ plus3pt%
\belowdisplayshortskip  7pt plus3.5pt minus0pt}
\def\small{\@setsize\small{13.6pt}\xipt\@xipt
\abovedisplayskip 13pt plus3pt minus3pt%
\belowdisplayskip \abovedisplayskip
\abovedisplayshortskip  \z@ plus3pt%
\belowdisplayshortskip  7pt plus3.5pt minus0pt
\def\@listi{\parsep 4.5pt plus 2pt minus 1pt
            \itemsep \parsep
            \topsep 9pt plus 3pt minus 3pt}}

\def\underline#1{\relax\ifmmode\@@underline#1\else
        $\@@underline{\hbox{#1}}$\relax\fi}
\@twosidetrue
\relax

\catcode`@=12

\catcode`\@=11

\def\section{\@startsection{section}{1}{\z@}{3.5ex plus 1ex minus
   .2ex}{2.3ex plus .2ex}{\large\bf}}

\def\ps@headings{\def\@oddfoot{}\def\@evenfoot{}
\def\@oddhead{\hbox{}\hfill
        \makebox[.5\textwidth]{\raggedright\ignorespaces --\thepage{}--
        \hfill }}
\def\@evenhead{\@oddhead}
\def\subsectionmark##1{\markboth{##1}{}}
}

\ps@headings

\catcode`\@=12

\relax

\def\figcap{\section*{Figure Captions\markboth
        {FIGURECAPTIONS}{FIGURECAPTIONS}}\list
        {Fig. \arabic{enumi}:\hfill}{\settowidth\labelwidth{Fig. 999:}
        \leftmargin\labelwidth
        \advance\leftmargin\labelsep\usecounter{enumi}}}
 \relax
\def\tablecap{\section*{Table Captions\markboth
        {TABLECAPTIONS}{TABLECAPTIONS}}\list
        {Table \arabic{enumi}:\hfill}{\settowidth\labelwidth{Table 999:}
        \leftmargin\labelwidth
        \advance\leftmargin\labelsep\usecounter{enumi}}}
 \relax
\def\reflist{\section*{References\markboth
        {REFLIST}{REFLIST}}\list
        {[\arabic{enumi}]\hfill}{\settowidth\labelwidth{[999]}
        \leftmargin\labelwidth
        \advance\leftmargin\labelsep\usecounter{enumi}}}
 \relax

\catcode`\@=11

\def\ps@headings{\def\@oddfoot{}\def\@evenfoot{}
\def\@oddhead{\hbox{}\hfill
        \makebox[.5\textwidth]{\raggedright\ignorespaces --\thepage{}--
        \hfill }}
\def\@evenhead{\@oddhead}
\def\subsectionmark##1{\markboth{##1}{}}
}

\ps@headings

\relax

\def\firstpage#1#2#3#4#5#6{
\begin{document}

\begin{titlepage}
\nopagebreak
\title{\begin{flushright}
      \vspace*{-1.3in}
        {\normalsize  SUSX-TH-98-01 }\\[-5mm]
        {\normalsize hep-th/9802099}\\[-5mm]
{\normalsize February 1998}\\[.5cm]
\end{flushright}
\vspace{10mm}
{\large \bf #3}}
\vspace{1cm}
\author{\large #4 \\ #5}
\maketitle
\vskip -1mm
\nopagebreak
\begin{abstract}
{\noindent #6}
\end{abstract}
\vspace{1cm}
\begin{flushleft}
\rule{16.1cm}{0.2mm}\\[-3mm]
$^{1}${\small E--mail: c.kokorelis@sussex.ac.uk}
\end{flushleft}
\thispagestyle{empty}
\end{titlepage}}
\newcommand{\dal}{\raisebox{0.085cm}
{\fbox{\rule{0cm}{0.07cm}\,}}}
\newcommand{\bb}{\begin{eqnarray}}
\newcommand{\ee}{\end{eqnarray}}
\newcommand{\p}{\partial}
\newcommand{\bp}{{\bar \p}}
\newcommand{\bR}{{\bf R}}
\newcommand{\bC}{{\bf C}}
\newcommand{\bZ}{{\bf Z}}
\newcommand{\bS}{{\bar S}}
\newcommand{\bT}{{\bar T}}
\newcommand{\bU}{{\bar U}}
\newcommand{\bA}{{\bar A}}
\newcommand{\bh}{{\bar h}}
\newcommand{\bu}{{\bf{u}}}
\newcommand{\bv}{{\bf{v}}}
\newcommand{\D}{{\cal D}}
\newcommand{\s}{\sigma}
\newcommand{\Sg}{\Sigma}
\newcommand{\ket}[1]{|#1 \rangle}
\newcommand{\bra}[1]{\langle #1|}
\newcommand{\non}{\nonumber}
\newcommand{\ph}{\varphi}
\newcommand{\la}{\lambda}
\newcommand{\ga}{\gamma}
\newcommand{\ka}{\kappa}
\newcommand{\m}{\mu}
\newcommand{\n}{\nu}
\newcommand{\th}{\vartheta}
\newcommand{\Lie}[1]{{\cal L}_{#1}}
\newcommand{\eps}{\epsilon}
\newcommand{\bz}{\bar{z}}
\newcommand{\bX}{\bar{X}}
\newcommand{\om}{\omega}
\newcommand{\Om}{\Omega}
\newcommand{\we}{\wedge}
\newcommand{\La}{\Lambda}
\newcommand{\bOm}{{\bar \Omega}}
\newcommand{\CA}{{\cal A}}
\newcommand{\CF}{{\cal F}}
\newcommand{\CbF}{\bar{\CF}}
\newcommand{\CAM}{\CA^{(M)}}
\newcommand{\CAS}{\CA^{(\Sg)}}
\newcommand{\CFS}{\CF^{(\Sg)}}
\newcommand{\I}{{\cal I}}
\newcommand{\al}{\alpha}
\newcommand{\be}{\beta}
\newcommand{\cm}{Commun.\ Math.\ Phys.~}
\newcommand{\pr}{Phys.\ Rev.\ D~}
\newcommand{\pl}{Phys.\ Lett.\ B~}
\newcommand{\ibar}{\bar{\imath}}
\newcommand{\jbar}{\bar{\jmath}}
\newcommand{\np}{Nucl.\ Phys.\ B~}
\newcommand{\e}{{\rm e}}
\newcommand{\gsi}{\,\raisebox{-0.13cm}{$\stackrel{\textstyle
>}{\textstyle\sim}$}\,}
\newcommand{\lsi}{\,\raisebox{-0.13cm}{$\stackrel{\textstyle
<}{\textstyle\sim}$}\,}
\date{}
\firstpage{95/XX}{3122}
{
The Master Equation for the Prepotential} 
{ Christos Kokorelis$^{1}$}
{\normalsize
Centre of Theoretical Physics\\
\normalsize Sussex University\\
\normalsize Falmer, Brighton, BN1 9QH, United Kingdom}
{ 
The perturbative prepotential and the K\"ahler metric of the vector
multiplets of the N=2 effective low-energy heterotic strings is
calculated directly in $N=1$ six-dimensional toroidal
compactifications of the heterotic string vacua.
This method provides the solution for the one loop correction to the
N=2 vector multiplet prepotential for compactifications of
the heterotic string
for any rank three and four models,
as well for compactifications on $K_3 \times T^2$.
In addition, we complete previous calculations, derived from string
amplitudes, by
deriving the differential equation for the third
derivative of the prepotential with respect of the usual
complex structure U moduli of the $T^2$ torus. 
Moreover, we calculate the one loop prepotential, using its modular 
properties, for N=2
compactifications of the heterotic string
exhibiting modular groups similar with those
appearing in N=2 sectors of N=1 orbifolds based on non-decomposable 
torus lattices and on N=2 supersymmetric Yang-Mills. }

\newpage

\section{Introduction}





One of the most important aspects of string dualities involve
comparisons of the effective actions between 
$N=2$ compactifications of the ten dimensional heterotic string to
lower dimensions and type II superstrings.
A key future for testing these dualities 
is the use of prepotential, which describes the $N=2$ effective 
low energy theory of vector multiplets in a general supergravity theory.

Guidance in working, with the vector multiplet heterotic prepotential at the
string theory level, comes form similar results 
from $N=2$ supersymmetric Yang-Mills\cite{seiwi1}.
At the level of $N=2$ supersymmetric $SU(r+1)$ Yang-Mills the
quantum moduli space was associated\cite{club} with a particular genus r
Riemann surface parametrized by r complex moduli and 2r periods
$(\alpha_{D_{i}},\alpha)$.
When matter is not present it allows for generic values 
of the scalar field of the theory to be broken down to the 
Cartran sub-algebra and it is described from r $N=2$ abelian vector 
supermultiplets.
The theory is dominated from the behaviour of the holomorphic 
function $\cal F(\cal A)$, namely the prepotential.
The supersymmetric non-linear $\sigma$-model is described
by the K\"ahler potential 
$K(A,{\bar A})=Im\{\frac{\partial {\cal F}(A)}{\partial A} {\bar A}\}$,
while the metric in its moduli space 
$Im(\tau(A)) = Im\{{\partial^2 {\cal F}/{\partial A}^2}\}$ is
connected to the complexified variable $\theta_{eff}/{\pi} + 
{8\pi i}(g_{eff}^{-2})\equiv \tau(A)$. The metric
is connected\footnote{$\alpha$ is the 
scalar component of the superfield $\cal A$ and can play the role 
of the Higgs field.}to the interpretation of the periods 
$\cal \pi$
\begin{eqnarray}        
{\cal \pi}=\left(\begin{array}{c}{\alpha}^{i}_D\\{\alpha^i}
\end{array}
\right),\;\;\;\;\alpha^i_D =\frac{\partial{\cal F}}{\partial
{\alpha}^i},\;i=1,\dots,r
\label{period1}                                                                                                      
\end{eqnarray}
as an appropriate family of a meromorphic one-forms associated
with $\lambda$, 
\begin{equation}
{\alpha}^i_D = \oint_{\alpha_i} \lambda,\;\;\;{\alpha}^i =
\oint_{\beta_i}\lambda,\;\;\tau =\frac{{d{\alpha}^i_D}/du}{
d{\alpha_i}/du}. 
\label{curve3} 
\end{equation}
Here, ${\alpha}_i, \beta_i$ form a basis\footnote{              
The cycles $\alpha, \beta$ form a basis of the first homology group 
$H_1(E_g, Z) = Z^{2g}$, where $E_g$ a Riemann surface at 
genus $g$. The intersection of cycles in the canonical basis 
takes the form $(a_i, b_j) = - (b_j,a_i)=\delta_{ij}$.}
of the homology cycles of the hyperelliptic  curve 
which has the same moduli space as $N=2$
supersymmetric Yang-Mills theory. 
This means that the metric on the moduli space of supersymmetric 
Yang-Mills is identified with the period matrix of the hyperelliptic
curve.
We have to notice that the gauge group is always abelian
and the classical enhancement symmetry point 
is absent from the theory. Only the weak coupling point
is included in the theory, making perturbative calculations for 
supersymmetric Yang-Mills reliable only in this area of moduli 
space.
The general picture emerging from the study of the 
supersymmetric Yang-Mills is that the vacuum expectation 
values of the Higgs fields break the theory to its maximal abelian 
subgroup. 
In addition, the exact theory is always in the Higgs phase 
missing the perturbative point where the non-abelian gauge symmetry
is restored.

Lets us now return in string theory to see, how the picture of
supersymmetric Yang-Mills is modified and moreover, its connection
to the prepotential and the one loop K\"ahler metric
in $N=2$ compactifications of the heterotic string.
Various equivalences between the different 
theories have been proposed and the picture emerging is that
the different string theories are expansions of a
more fundamental theory around different points in 
the moduli space of string vacua.
We mention string-string duality, where type IIA compactified on
$K_3$ manifold with $N=4$ supersymmetry has the same moduli space as the
heterotic string
on a $T^4$ torus\cite{huto,sen1,var,hastro,asmo,aspo1} with $N=4$ 
supersymmetry. Now the strong coupling of the heterotic string is mapped
into the weak coupling of IIA. 
For $N=2$ type IIB the string generalization of Seiberg-Witten's(SW)
quantum theory is provided by the conifold\footnote{
The exact effective low energy coupling constant for this theory
have been calculated using mirror symmetry.} transitions
of wrapped three-branes\footnote{The three brane of the type IIB can
end\cite{stroo} on the RR
five-brane.} on Calabi-Yau spaces.
Type IIB in ten dimensions admits extremal black holes  
solutions in the RR sector of the theory. They represent BPS saturated 
p-brane solitons.   
Compactification of type IIB on a Calabi-Yau space produces 
$h^{(1,1)} + 1$ supermultiplet moduli with +1 associated with the 
dilaton and 
$h^{(2,1)}$ vector multiplets.
In addition, it gives the abelian gauge group $U(1)^{h^{(2,1)} + 1}$.
In general special geometry\footnote{See next section.}
applied to the compactification of type IIB on the 
Calabi-Yau space in four dimensions,
requires that the scalar component Z   
and the prepotential F of the vector multiplets to be given by
the period of the three form $\Omega$ over the canonical
homology cycles ${a_I}, b_I$ as 
\begin{eqnarray}
Z_I = \int_{a_I} \Omega,\;\;\;\frac{\partial F}{\partial Z_I}= 
\int_{b_I}\Omega,\;\;\;I=1,\dots,h^{(2,1)}.
\label{kotera1}
\end{eqnarray}
Here, $\Omega$ is the holomorphic three form describing the 
complex structure of the Calabi-Yau space.
BPS states are $\propto |-\hat{\nu}_e^I Z_I + \hat{\nu}_m^I
F_I|$. The integers $\hat{\nu}_e, \hat{\nu}_m$ are the 
electric and magnetic charges of the threebrane wrapped around
the three surfaces ${a_I}, b_I$.
The appearance of a logarithmic singularity in the 
K\"{a}hler metric at the conifold point $Z{=}0$, involved in 
the compactification of type IIB on the Calabi-Yau space, is then 
identified\cite{stro1} with the extremal three brane black hole 
becoming massless.  In analogy with SW theory, 
the three-brane becomes massless when the associated cycles 
vanish. 
The appearance of the singularity, when the corresponding
3-cycles along the 3-surfaces vanish, is then identified with
the existence\cite{stro1} of a massless black hole solution in the
metric of type IIB for the 3-brane.
However now, the singularity is not a weak coupling singularity
but a strong coupling singularity.
This is the analog of   
Seiberg-Witten appearance of the massless monopole singularity
in type IIB.
Here we have to remind that study of heterotic string theory,
analog of Seiberg-Witten theory, nessessarily involves 
the calculation of the one loop prepotential and the one loop  
K\"ahler metric since effective gauge couplings in N=2
supergravity\cite{wkll} depend on the
moduli of the vector multiplets only.
Later, in section five, we will derive the
general equation describing the prepotential of $N=2$ compactifications
of a heterotic rank three model which has a dual pair, in the sence
of \cite{kava}, described by a IIB model compactified on a Calabi-Yau.

In this work we are interested, among the web of dualities, only in the 
proposal of \cite{kava,fhsv} which
provided evidence for the exact nonperturbative equivalence
of dual pairs, the 
heterotic string compactified on $K_3 \times T_2$, with 
IIA compactified on a Calabi-Yau threefold.
The proposal identifies the moduli spaces of heterotic string and its 
dual IIA as ${\cal M}_V^{heterotic}= 
{\cal M}_V^{IIA}$ and ${\cal M}_H^{heterotic}={\cal M}_H^{IIA}$, where
the subscripts refer to vector multiplets and hypermultiplets 
respectively. 
In other words, for models which are dual pairs,
the exact prepotential for the vector multiplets, including perturbative and
non-perturbative corrections is calculated from the IIA side, where the
tree level result is exact.
On the contrary, if we want to calculate the exact hypermultiplet
prepotential
this is calculated from the heterotic string on $K_3 \times T^2$.
This is exactly Strominger's proposal that the absence of
neutral perturbative couplings between vector multiplets and
hypermultiplets survives nonperturbative string effects.
In this sence the complete prepotentials for the vector
multiplets for the two "different" theories match, including perturbative
and non-perturbative corrections, and ${\cal F}^{het}= {\cal F}^{IIA}$.
Compactification
of the heterotic string on $K_3 \times T_2$ and of type IIA
on a Calabi-Yau threefold produce models
with N=2 supersymmetry in four dimensions.
These models have been tested to be dual to each other,
at the level of effective theories, and for low numbers of
$h^{(1,1)}$ vector multiplet moduli\cite{kava}.
By using mirror symmetry, 
 we can map the vector multiplet sector of IIA to its mirror
type IIB $h^{(2,1)}$ complex structure moduli space\footnote{
Application of the mirror symmetry on the type IIB
interchanges three-cycles with two cycles and three-branes with 
two branes.}.
The procedure then concentrates in the comparison of the 
complex structure moduli\cite{liya} effective theories, of type IIB,
with 
$N=2$ heterotic compactifications of rank three or four 
models\cite{kava,anpa}.
In this sence, in type IIB the result for the effective gauge coupling 
$\propto \log(Z) $ is exact 
to all orders of perturbation theory, since the dilaton in type IIB 
belongs to the hypermultiplet sector and are no couplings allowed
between\cite{DWLVP} vector multiplets and hypermultiplets.

The singularity structure of the Calabi-Yau can help us to 
produce transitions between different vacua by changing the
Hodge numbers of the manifold.  
Note, that for a Calabi-Yau space 
exploring the Higgs transitions at the conifold singularities 
can produce extremal transitions between different 
Calabi-Yau vacua\cite{stro2,cande,cande1}.
Exploring singularity transitions in heterotic string compactified
in $K_3 \times T_2$, can produce dual models, whose Hodge numbers 
match known type II models on Calabi-Yau three folds.  
This was confirmed with the 
construction 
of such duals in \cite{afiq1}.

In section four, we will calculate the one loop correction to the  
perturbative prepotential of the vector multiplets
for the heterotic string compactified on a six dimensional orbifold.
It comes from the general solution from the one loop K\"ahler 
metric\cite{afgnt,anto3}.
The one loop correction to the perturbative prepotential has already
been calculated before in \cite{afgnt} from string amplitudes. 
Our procedure is complementary to \cite{afgnt} since we calculate, 
contrary to \cite{afgnt} where the third derivative of the prepotential 
with respect to the T moduli was calculated, the third derivative of the 
prepotential with respect to the U moduli. 
The logarithmic singularity, consequence of the gauge symmetry enhancement 
at a specific point in the moduli space do appear in the prepotential
and the K\"ahler metric.

Furthermore, we {\em establish a general procedure} for calculating
one loop corrections to the one loop prepotential, not only
for $N=1$ six dimensional heterotic strings toroidally compactified on 
four dimensions,
which has important implications for any compactification of the 
heterotic string having(or not) a type II dual. This procedure is an 
alternative way to the calculation of the prepotential which was 
performed indirectly in \cite{mouha} via the effective gauge 
couplings.

In this work we will continue the 
work of\cite{anto3,wkll,afgnt}.
We will calculate directly the integral representation
of $N=2$                                                         
 vector multiplet prepotential of toroidal
compactifications of the 
 heterotic string\footnote{However, due to factorization
properties of the $T^2$ subspace of the heterotic Narain lattice
and instanton embedding independence into the gauge bundle over $K_3$,
the same result can be applied to any heterotic string 
compactification on $K_3 \times T^2$ for any rank four
models.}.
The one loop K\"{a}hler metric in the moduli space of 
vector multiplets of toroidal compactifications of $N=1$ six dimensional
orbifold\cite{Seiberg}
compactifications\cite{walt,erle} of the heterotic string follows directly
from this result.

In section two we will describe general properties of $N=2$ heterotic 
strings. In addition, we describe properties of the moduli space of
compactification of 
the heterotic string on a $K_3 \times T_2$ manifold.
In addition, we describe
elements of the special K\"ahler geometry 
describing
$N=2$ locally sypersymmetric theory of the heterotic string 
with emphasis on the couplings
of vector multiplets. 

In section three we will 
describe our results for the case of toroidal compactifications
of $N{=}1$ six dimensional
orbifold
compactification of the heterotic string, where the underlying
torus lattice does not "decompose" as $T^2 \oplus T^4$, the $T^4$ could be 
an orbifold limit of $K_3$.
The moduli of the unrotated complex plane, e.g $T^2$ with 
a shift,  
 has a modular symmetry
group that is a subgroup of $SL(2,Z)$. In particular, we consider
this modular symmetry, e.g $\Gamma_0(3)$,
to be one of those appearing in $N=2$ sectors non-decomposable $N=1$ orbifold
compactifications\cite{erkl,deko}.

In section four
we describe our results for the one loop prepotential of
toroidal compactification of the $N=1$
six dimensional heterotic
string to four dimensions. 
In the description of calculating the prepotential of vector
multiplets from string amplitudes
we will follow the work of \cite{afgnt,anto3}.
We use string amplitudes 
to calculate directly the one loop prepotential via its relation
to the one loop K\"ahler metric.
The calculation comes from the use of string amplitudes
of \cite{anto3,afgnt}.  Automatically this calculates
one-loop
corrections to the K\"ahler metric for the moduli of 
the usual vector multiplet T, U moduli fields of the $T^2$ torus
appearing in $N=2$ $(4,4)$ compactifications of the heterotic 
string. The calculation
on the quantum moduli space takes into consideration points of 
enhanced gauge symmetry.

In section five we describe our results for
the $N=2$ prepotential of any rank three compactifications of
the heterotic string.
 A general result\cite{fava1} concerning the geometry, local issues,
behind the existence of heterotic duals, is that
the Calabi-Yau manifold in the IIA side can be written  as
a fibre bundle with base\footnote{
 $P^1$ is the complex projective space with homogeneous coordinates
$[x_o, x_1]$.}$P^1$ and generic fiber the $K_3$ surface\cite{klm1},
while the dual heterotic, on $K_3 \times T^2$, could be written
instead as a bundle with base $P^1$ and generic fibre $T^4$.
Existence of the IIA dual in the Calabi-Yau threefold phase,
as a global issue,
with the dual heterotic string admitting a weekly coupled phase while
the dual type IIA realization is in the strongly coupled phase,
can happen only when\cite{lla} the generic fibre is
the $K_3$ surface and the base is $P^1$.
In section five we apply the {\em general procedure} for calculating 
the one loop 
heterotic prepotential in particular dual type II models compactified on
Calabi-Yau,
 equivalent to rank three heterotic models\cite{osa} at a specific
limit.
The Calabi-Yau models incorporate the
 $K_3$ fiber
 structure \cite{klm1}
of the type II dual realization.


\section{Properties  of $N=2$ heterotic vacua}

{\em Generalities}

In the next section we will calculate the perturbative prepotential using
its modular properties for a class of models, possible orbifold limits
of $K_3$, whose modular group is similar to those appearing  
$N=2$ sectors of $N=1$ $(2,2)$ symmetric non-decomposable orbifolds.
The list of $(2,2)$ vacua includes the Calabi-Yau
compactifications \cite{chsw}, orbifolds \cite{dhvw,imnq}, tensor
products of minimal models \cite{ge1} or generalizations \cite{kasu}.
For abelian $(2,2)$ orbifolds constructed
by twisting a six dimensional torus, the point group rotation
is accompanied by a similar rotation in the gauge degrees of freedom.
The four dimensional gauge group in this case is enlarged beyond
$G= E_6 \otimes E_8$ by a factor that can be a $U(1)^{2},\;SU(2) 
\times U(1)$, if $P=Z_4$ or $Z_6$, or $SU(3)$ if $P=Z_3$.
If we symbolize by $h^{(1,1)}$ the number\footnote{For 
compactifications on Calabi-Yau manifolds, $h^{(1,1)}$ and $h^{(2,1)}$ 
represent the Hodge numbers of the manifold.}
of $(1,1)$ moduli in the
untwisted sector then we have respectively  
$h^{(1,1)}= 3, 5$ and $9$.
Twisted moduli are not neutral with respect to G and are not moduli 
of the orbifold.
On the other hand 
$N=1$ orbifold compactifications of the ten dimensional heterotic string
have, in four dimensions, $N=1$ and $N=2$ sectors twisted
sectors.
The geometry of those $N=2$ sectors is described 
from special geometry, which will be desribed in the following.
Abelian $(2,2)$ orbifolds can flow to a Calabi-Yau vacuum, by
blowing up the twisted moduli fields, giving them vacuum expectation
values\cite{HV}.  The $N=2$ sectors exhibit modular groups 
which are similar to those appearing to $N=2$ supersymmetric Yang-Mills 
with number of hypermultiplets zero or two\cite{klemmos,nahm}.

Of particular interest in this work is the $N=2$ toroidal 
compactifications of the six dimensional heterotic vacua.
The moduli of the two torus is parametrized from the relations
$T=2(B+i\sqrt{G})$ and $U=1/G_{11}(G_{12} + \sqrt{G})$, where $G_{ij}$
is the lattice metric of the $T^2$, $\sqrt{G}$ is determinant and B
the constant antisymmetric tensor background field. The moduli T, U 
represent the scalar components of two $N=2$ vector multiplet fields
in four dimensions.

At the classical level, moduli are described by a flat 
potential\cite{Di1,abak,illt} to all orders of perturbation theory.
Compactifications of the heterotic string in four dimensions with
$N=2$ supersymmetry involve a $U(1) \times U(1)$ gauge group
from the untwisted $T^2$ subspace.
The T, U moduli subspace, exhibits an $SL(2,Z)_T \times SL(2,Z)_U
\times Z_2^{T \leftrightarrow U}$
classical duality group, 
and corresponds to the coset 
space ${\frac{SO(2,2)}{SO(2) \times SO(2)}}|_{T,U}$. 
The same type of moduli appears when we further compactify\cite{walt,mouha},
the
$N=1$ six dimensional heterotic string compactified on the manifold $K_3$,
on a two torus.
The gauge group associated with the two torus can be futher enhanced
at
special points in the moduli space, namely the $T=U$ line   
the gauge group becomes enhanced to $SU(2) \times U(1)$.
It can become enhanced to $SO(4)$ or $SU(3)$ along the
$T=U=i$ or $T=U= e^{{2\pi i}/3}$ lines respectively.
In general the
classical moduli space, of r vector multiplets in the $T^2$
subspace, is the group\cite{special} 
\begin{equation}
\left. \frac{SU(1,1)}{U(1)}\right|_{\makebox{dilaton}} \left. \times 
\frac{O(2,r)}{O(2) \times O(r)}\right/ O(2,r;Z),
\label{mpam}
\end{equation}
where the first factor corresponds to the complex dilaton D

 S.
For our case, the classical duality group comes with $r=2$.
Here, $O(2,2;Z)$ is the target space duality group. The theory enjoy 
the non-trivial global invariance
i.e identifications  under target space duality symmetries
\cite{DUAL,dis} the
$PSL(2, {\bf Z})_{T} \times PSL(2, {\bf Z})_{U}$ dualities acting as
\begin{equation}
T\rightarrow {aT-ib\over icT+d},\;\;\;\;\;
U\rightarrow{a'U-ib'\over ic'U+d'}.
\label{brum}
\end{equation}

The same vector multiplet, as eqn.(\ref{mpam}) appears 
for generic $(4,4)$ 
compactifications 
of the heterotic string on the $K_3 \times T^2$.
For compactifications on $K_3$ manifolds, the moduli space of metrics
with $SU(2)$ holonomy associated 
to complex and K\"ahler deformations is 
$\textstyle{{\cal M}=\frac{SO(19,3)}{SO(19) \times SO(3) \times
SO(19,3;Z)} \times R^{+}}$, where $R^{+}$ is  
associated\cite{asmo} with the volume of $K_3$.
Adding the moduli coming from deformations of the antisymmetric
tensor we get the moduli space of hypermultiplets\cite{NS} of $K_3$ 
$\textstyle{\frac{SO(20,4)}{ SO(20) \times SO(4,20;Z) \times SO(4)}}$.

Lets us now describe some properties\footnote{For a
review of $N=2$ heterotic strings see \cite{wkll}.}
 of the low 
energy
effective actions of $N=2$ effective string theories.
In $N=2$ supersymmetric Yang-Mills theory the action for the vector
multiplets is described
by a holomorphic prepotential ${\tilde F}(X)$, where $X^A$ ($A=1,\ldots,n$)
are the complex scalar components of the corresponding vector
superfields. 
The couplings of the classical vector multiplets with supergravity 
are
determined by another holomorphic function $F(X)$, the prepotential function 
which is a holomorphic function of
$n+1$ complex variables $X^I$ ($I=0,1,\ldots,n$) and 
it is a homogeneous function of degree 
two\cite{DWVP} in the fields  $X^I$ .
However, in $N=2$ supergravity theories, supersymmetry demands an additional 
vector
superfield $X^0$ which account for the accommodation of the 
graviphoton.
It stands for the $\textstyle{I=0}$ component of the vector multiplets and it
belongs to a compensating multiplet.
The graviphoton is the vector component of the compensating 
multiplet and is the spin one gauge boson of the 
$N=2$ supergravity multiplet.
The coordinate space of physical scalar fields
belonging to vector multiplets of an $N=2$
supergravity\cite{DWVP,sukou} is described from special K\"ahler 
geometry \cite{special},
with the K\"ahler metric
$\textstyle{g_{AB}=\partial_A \partial_{\bar B} K(z,\bar z)}$
resulting from a K\"ahler potential of the form
\begin{equation}
K(z,\bar z)=-\log\Big(i{\bar X}^I(\bar z)\,F_I(X(z) -iX^I(z)
\bar F_I({\bar X}({\bar z})\Big),\; F_I=\frac{\partial F}{\partial X_I} 
,\;{\bar F}_I =\frac{\partial F}{\partial {\bar X}_I}.
\label{KP}
\end{equation}
The spectrum of $N=2$ heterotic strings in four dimensions
contains, among other fields, the 
dilaton D and the antisymmetric tensor\footnote{connected
to the axion via a duality transformation}
and the graviton.
The axion is subject to the discrete Peccei-Quinn symmetry
to all orders of perturbation theory. Since the axion is connected 
through a duality 
transformation to the antisymmetric tensor field, whose
vertex operator decouples at zero momentum, this means that
every physical amplitude involving $B_{\mu \nu}$ at zero momentum 
is zero.
As a result the effective theory of the heterotic superstring
is independent of the field $B_{\mu \nu}$ at zero momentum and the 
coupling of 
field appear only through its derivative. 
The dilaton and the axion belong to a vector multiplet.
Since the axion couples to the dilaton D via the complex scalar
S, which we will refer next as the dilaton, 
we conclude that any dependence
of holomorphic quantities, e.g the Wilsonian gauge couplings,
will be through the combination $S+{\bar S}$.
However, these arguments
are not valid non-perturbatively.
As a result the non-perturbative corrections to the prepotential
of $N=2$ compactifications of the heterotic string depend on S.
However, they will be of no particular interest to us,
since we will discuss the one loop correction to the prepotential
at the semiclassical limt $S \rightarrow \infty$.

The prepotential
for the classical moduli space of vector multiplets reads
\begin{equation}
F(X)= -{X^1\over X^0}\Big[X^2X^3-\sum_{I = 4}^n(X^{I})^2\Big]\,.
\label{ourF}
\end{equation}
while the values of the moduli are identified as 
\begin{equation}
S=-i {X^1\over X^0}\,,\quad T=-i {X^2\over X^0}\,,\quad
U=-i {X^3\over X^0}\,,\quad \phi^i=-i {X^{i+3}\over X^0}\ ,\quad
(i=1, \ldots, P)\ ,
\label{modid}
\end{equation}
with the remaining  $X^I,C^{a} = -i X^{a+P+3}/ X^0, a = p+4,\ldots,n$ 
correspond \cite{wkll,special} to matter scalars.
Matter scalars are members of the hypermultiplets while the 
scalars in the Cartran subalgebra of the non-abelian gauge group  of the 
heterotic vacuum can be used to break the gauge group.

From the values of the moduli previously given, it follows that the
the K\"ahler potential is
\begin{equation}
K= -\log \left( (S+\bar S) [(T+\bar T)(U+\bar U)-\sum_{i}
({\phi}^i+{\bar \phi}^i)^2 -\sum_{a}(C^a+{\bar C}^a)^2 \right).
\label{KKKK}
\end{equation}

\section{ One loop prepotential - perturbative aspects}

Since we have finish our discussion of the general properties of the  
$N=2$ heterotic strings, we will now discuss the calculation of 
perturbative corrections to the one loop prepotential.
We are interested on those heterotic strings which exhibit modular groups
similar to those appearing in the calculation of thresholds corrections
in non-decomposable\cite{erkl} $N=1$ symmetric orbifolds\cite{deko}.

Let us expand at the moment the expression of eqn.(\ref{ourF})  
around small values of the non moduli scalars $C_a$ as in
(\ref{KKKK}) 
\begin{equation}
F = -S(TU - \sum_i \phi^i \phi^i) + {h}(T,U,\phi^i),\;\;
\label{hfexp}
\end{equation}
or
\begin{eqnarray}
F = -d_{sij}T^i T^j S  +  {h}(T,U,\phi^i),\;\; d_{sij}= diag(+,-,\dots,-)
\nonumber\\
T^1=T,\;T^2=U,\;T^i = \phi^i,\;i \neq 1,2.
\label{hfexp2}
\end{eqnarray} 
The function ${h}$, the one loop correction to the 
perturbative prepotential,
enjoy a non-renormalization theorem, namely it receives perturbative
corrections only up to one loop order. Its higher loop corrections,
in terms of the $1/{(S+ \bar S)}$, vanish
due to the surviving of the discrete Peccei-Quinn 
symmetry to all orders of perturbation theory as a quantum symmetry.

In that case, under target space duality 
\begin{equation}
T \stackrel{SL(2,Z)_T}{\rightarrow} \frac{aT-ib}{icT+d},\;\; U 
\stackrel{SL(2,Z)_T}{\rightarrow} U,
\label{tathey}
\end{equation}
we get
\begin{equation}
h(T,U) \stackrel{SL(2,Z)_T}{\rightarrow} \frac{h(T,U)+ \Xi(T,U)}{(icT+d)^2}
\label{Ton1h}
\end{equation}
and a similar set of transformations under $PSL(2,Z)_{U}$. The net
result is that ${\partial}_{T}^{3} h^{(1)}(T,U)$ is a singled valued 
function of weight $-2$ under U-duality and $4$ under T-duality. 
The prepotential h modifies the K\"ahler 
potential, in the lowest order of expansion in
the matter fields, as
\begin{equation}
K = - \log[(S +{\bar S}) + V_{GS}] -
\log(T + {\bar T})^2 -\log(U + {\bar U})^2,
\label{Tontinar}
\end{equation}
where 
\begin{equation}
V_{GS} = \frac{ 2 ( h + {\bar h}) - ( T + {\bar T})(\partial_T h +
\partial_{\bar T} {\bar  h}) - ( U + {\bar U} )( \partial_U h +
\partial_{\bar U} {\bar h} ) }{(T + {\bar T}) ( U + {\bar U}) }
\label{klroupa1}
\end{equation}
is the Green-Schwarz term\cite{dere} which contains\footnote{
In the next section, we will see practically the direct calculation
of the $V_{GS}$ via the calculation of h. However, our notation will be
different following the spirit of \cite{afgnt} and h will be 
denoted by f.}  
the mixing of the dilaton with the moduli at one loop order.  
Rememebr, that for the conventions used at this section the
dilaton is defined as $ < S > = 4 \pi /g^2 - i \theta / 2 \pi$. 

The prepotential for $N=2$ orbifold
compactifications of the heterotic string\cite{walt,hemo,erkl}.
was calculated, from the use of its modular properties 
and singularity structure in \cite{wkll}.
Here, we adopt a similar approach
to calculate the prepotential of vector multiplets.
We discuss the calculation of 
the prepotential for the case where the moduli subspace of the 
Narain lattice associated with the T, U moduli exhibits a modular 
symmetry\cite{deko,blst1,blst2} group $\Gamma^o(3)_T \times \Gamma^o(3)_U$. 
The same modular symmetry group appears\cite{deko} in the $N=2$ sector
of the $N=1$ $(2,2)$ symmetric non-decomposable $Z_6$ orbifold defined 
on the lattice $SU(3) \times SO(8)$. In the third complex plane associated
with the square of the complex twist $(2,1,-3)/6$ the mass operator
for the untwisted subspace was given\footnote{We changed notation.
All moduli are rescaled by i.}
 to be
\begin{equation}
m^2= \sum_{m_1, m_2, n_1, n_2\;\in\;Z} \frac{1}{2T_2U_2}
|TU^{\prime}n_2 + T n_1 - U^{\prime}m_1 + 3m_2|^2,\;\;\;U^{\prime} = U+2.
\label{arga1}
\end{equation}
Let us forget the $N=1$ orbifold nature of the appearance
of this $N=2$ sector. Then its low energy supergravity theory 
is described by the underlying special geometry. 
The question now is if calculating the prepotential using its modular 
properties and the singularity structure, as this was calculated
for decomposable\footnote{We use this term
in connection with the same type of modular symmetries
appearing in $N=1$ decomposable orbifold compactifications of the
heterotic string\cite{dhvw}.}
 orbifold compactifications of the heterotic 
string\cite{wkll}, has any type II dual realization.
We believe that it is the case.
In the analysis of the map between type II
and heterotic dual supersymmetric string theories\cite{liya,klm1}  
it was shown that subgroups of the modular group do appear.
In particular some type II compactified on the Calabi-Yau three 
folds\cite{kava},
were shown\cite{klm1} to correspond 
in one modulus deformations of $K_3$ fibrations.
The modular symmetry groups appearing\cite{liya} are all connected
to the $\Gamma_o(N)_{+}$, the subgroup
of the $PSL(2,Z)$, the $\Gamma_o(N)$ group together
with the Atkin-lehner involutions $T\rightarrow \frac{-1}{NT}$.
We expect that the same prepotential, beyond describing the  
geometry of the $N=2$ sector of $Z_6$ in exact analogy to the  
decomposable case, may come form a compactification of the heterotic
string on the $K_3 \times T^2$. An argument that seems to give some support
to our conjecture was given in \cite{fava1}.
It was noted by Vafa and Witten 
that if we compactify a ten dimensional string theory on $T^2 \times X$,
where X any four manifold, acting with a $Z_2$ shift on the Narain lattice 
we get the modular symmetry group $\Gamma_o(2)_T \times \Gamma_o(2)_U$.
In this respect it is obvious that our calculation of the prepotential
may come from a shift
in a certain Narain lattice of $T^2$. We suspect that this is a $Z_3$
shift.
Furthermore, if we adopt N=2
conventions\cite{klemmos} in the study of dyon spectrum of $N=2$ 
supersymmetric Yang-Mills, the quantum symmetry groups $\Gamma_o(2)$,
$\Gamma^o(2)$ appear in $N_f=0, 2$ respectively with corresponding
monodromy groups $\Gamma^o(4)$,  $\Gamma_o(4)$.

From the mass operator (\ref{arga1}) we deduce that at the point $T=U$
in the moduli space of the $T^2$ torus of the untwisted plane, with
$n_1 = m_1=\pm 1$ and $n_2=m_2=0$, its $U(1) \times U(1)$ symmetry
becomes enhanced to $SU(2) \times U(1)$. Moreover, the third derivative
of the prepotential has to transform, in analogy to the 
$SL(2,Z)$ case, with modular weights -2 under ${\Gamma^o(3)}_U$ and
4 under $\Gamma^o(3)_T$ dualities.
Using the theory of modular forms requires, for the
calculation of the vector multiplet prepotential
of the effective N=2 low energy theory of heterotic strings,
 the analog of $SL(2,Z)$ j-invariant, for $\Gamma^o (3)$,
 the hauptmodul function.
This quantity is given by ${\omega(T)}$, where ${\omega(T)}$ is given 
explicitly by
\begin{equation}
\omega(T) = (\frac{\eta(T/3)}{\eta(T)})^{12}
\end{equation}
and represents the hauptmodul
for $\Gamma^0(3)$,
the analogue of $j$ invariant for $SL(2,\bf Z)$.
It is obviously automorphic under $\Gamma^0(3)$ and possess\footnote{We
would lile to thank D. Zagier for pointing this to us.}
a double pole at infinity and a double zero at zero.
It is holomorphic\cite{shoe} in the upper complex plane and at
the points
zero and infinity has the expansions
\begin{equation}
{\omega(T)}= t_{\infty}^{-1} {\sum}^{\infty}_{{\lambda}=0}
a_{\lambda} t^{\lambda}_{\infty}\;, {a_{o} \neq 0},\;\;
{\omega(T)}= t_{o}^{-1} {\sum}^{\infty}_{{\lambda}=0}
b_{\lambda} t^{\lambda}_{o}\;, {b_{o}\neq 0}
\end{equation}                                                          
at ${\infty}$ and $0$ respectively with $ t = e^{- 2\pi T}$.

In full generality, the hauptmodul
functions for the $\Gamma^0(p)$ are the functions\cite{apostol,kobl}
\begin{equation}
\Phi(\tau) = \left( \frac{\eta(\frac{\tau}{p})}{\eta(\tau)}\right)^r.
\label{haupt1}  
\end{equation}
 Here, $p= 2,3,5,7$ or 13 and $r=24/(p-1)$. For these values of p
the function in eqn.(\ref{haupt1}) remains modular invariant, i.e it
is a modular function.
The hauptmodul functions for
the group $\Gamma_0(p)$ are represented by 
\begin{equation}
(\frac{\eta({\tau})}{\eta(p\tau)})^r.
\label{hauptas1}
\end{equation}

The function $\omega^o(3)$ has a single zero at zero
and a single pole
at infinity. In addition, its first derivative
has a first order zero at zero, a pole at infinity and
a first order zero at $i{\sqrt{3}}$.
The modular form F of weight k of a given subgroup of the 
modular group $PSL(2,Z)=SL(2,Z)/{Z_2}$ is calculated 
from the formula 
\begin{equation}
\sum_{p \neq 0, \infty }{\nu_p} + \sum_{p=0,\infty} (width)
\times (order\;of\;the\;point) = \frac{\mu k}{12}.
\label{eksiso1}
\end{equation}
Here, $\nu_p$ the order of the function F, the lowest power in the  
Laurent expansion of F at p. The index $\mu$ for $\Gamma^o(3)$
is calculated from the expression\cite{shoe}
\begin{equation}
[\Gamma: \Gamma_o(N)] =N {\displaystyle\mathop{\Pi_{p/N}}}\;(1+p^{-1})
\label{extresi1}
\end{equation}
equal to four.
The width at infinity is defined as the smallest integer
such as the transformation $z \rightarrow ( z + \alpha )$ is in the group,
where $\alpha \in Z$. 
The width at zero is coming by properly transforming the width at infinity
at zero.
For $\Gamma^o(3)$ the width at infinity is 3 and the width at zero is 1.
The holomorphic prepotential can be calculated easily if we examine its
seventh derivative. The seventh derivative has modular weight
12 in T and 4 in U. In addition, it has a sixth order pole at the 
$T=U$ point whose coefficient A has to be fixed in order to produce the
logarithmic singularity of the one loop prepotential.
As it was shown\cite{afgnt,wkll} the one loop 
prepotential as T approaches $U_g = \frac{aU + b}{cU + d}$, where 
g is an $SL(2,Z)$ element\footnote{The same argument works for the 
subgroups of the modular group, but now there are 
additional restrictions on the 
parameters of the modular transformations. }
\begin{equation}
f \propto -\frac{i}{\pi} {\{(cU+d)T -(aU + b)\}^2} \ln(T - U_g).
\label{opli11}
\end{equation}
The seventh derivative of the prepotential is calculated to be
\begin{equation}
f_{TTTTTTT} = A \frac{\omega(U)_U^3  \omega(U)^5 (\omega^{\prime}(U))^3}
{(\omega(U) -\omega(\sqrt{3}))^2 (\omega(U) -\omega(T))^6} X(T),
\label{teleios1}
\end{equation}
where $X(T)$ a meromorphic modular form with modular weight 12 in T.
The complete form of the prepotential is
\begin{eqnarray}
f_{TTTTTTT} = A \left(\frac{[\omega(U)_U^3  \omega(U)^5 
(\omega^{\prime}(U))^3}{
{(\omega(U) -\omega(\sqrt{3}))^2 [(\omega(U) -\omega(T))^6}]}\right)
\left(\frac{\omega(T)_T^6}{\omega^2(T)\{(\omega(T)-
\omega (\sqrt{3}))^4 \}}\right).
\label{teleios2}
\end{eqnarray}

The two groups $\Gamma^o(3)$ and $\Gamma_o(3)$ are  conjugate to each 
other. If S is the generator 
\begin{eqnarray}
S= \left(\begin{array}{cc}
0&-1\\
1&0
\end{array}\right),\;\;we\;have\;\Gamma^o(3)=S^{-1} \Gamma_o(3) S.
\label{eftasa1}
\end{eqnarray}
So any statement about modular functions on one group is a statement 
about the
other. We have just to replace everywhere $\omega(z)$ by $\omega(3z)$ 
to go from a
modular function from the $\Gamma^o(3)$ to the $\Gamma_o(3)$.
In other words, the results for the heterotic prepotential with
modular symmetry group $\Gamma^o(3)$ may well be describe the
prepotential to the conjugate modular theory.

We have calculated the prepotential of a heterotic string
with a $\Gamma^o(3)_T \times \Gamma^o(3)_T \times Z_2^{T 
\leftrightarrow U}$ classical duality group.
The same dependence on the T, U moduli and its modular symmetry group
appears in the $\Theta^2$, $N=2$, sector 
of the $Z_6$ orbifold defined by the action of the 
complex twist $\Theta= exp[\frac{2 \pi i}{6}(2,3,-1)]$
on the six dimensional      
lattice $SU(3) \times SO(8)$, namely the $Z_6$-IIb.

If our weakly coupled heterotic model has a 
non-perturbatively equivalent Calabi-Yau dual type IIA model
then it has to come from generic fibers of a $K_3$
fibration. The generic fibers can be seen from the 
non-zero entries of the
intersection numbers 
\footnote{See for example (\ref{koufme1})  
and for more details section five.}
\begin{equation}
(D_{sij})=diag(+,-,\dots,-,0,\dots,0).
\label{newlinw1}
\end{equation}
The 
zero entries correspond to singular fibers, fibers which degenerate
at points in the moduli space,  to non $K_3$ surfaces like a smooth 
manifold, and correspond to the heterotic side to strong  
coupling singularities\cite{witta,kainou} seen 
non-perturbatively\cite{aspigr}.
Because of the maximum number of $K_3$ moduli 20, the number of 
generic fibers is constrained to be less than 20.
We believe that the nature of the lattice twist
of non-decomposable orbifolds is such that its form when acting
on the $N=2$ planes may correspond to orbifold limits\footnote{
C. Kounnas has suggested to me that some of them might
correspond to freely acting orbifolds.} 
of $K_3$. 
In this phase, the $K_3$ surface can be
written as an orbifold of $T^4$. The fixed points of $T^4$ under the 
orbifold action are the singular limits of $K_3$ because the metric
on the fixed point develops singularities. The singularities 
of $K_3$ follow an ADE classification pattern. In fact, because
at the adiabatic limit\cite{fava1}, 
we can do even the reverse, we can map the type II phase to the heterotic 
one. In the limit where the base of the fibration has a large area, but 
the volume of the $K_3$ fiber is of order one, we can replace the 
$K_3$ fibers in the heterotic side with $T^4$ fibers over a $P^1$
base.
In this form, the 
heterotic string description is replaced by $T^4$
fibers, namely the Narain lattice $\Gamma^{20,4}$.

\section{ One loop correction to the prepotential from string
amplitudes}
In the section (4.1) we review general properties involved in the
calculation of the one loop K\"ahler metric and the one loop
prepotential f. Then we discuss properties of the equation for $f_{UUU}$
and finally we introduce the equation for the rank four 
$N=2$ heterotic string compactifications.

\subsection{One loop contribution to the Prepotential/ K\"{a}hler metric}

{\em Preliminaries - Rank four models}

The one-loop
K\"{a}hler metric for orbifold compactifications of the heterotic 
string,
where the internal six torus decomposes into $T^2 \oplus T^4$, was 
calculated in \cite{anto3}. In this section, we will use the general 
form of the 
solution for the one loop K\"{a}hler metric appearing in
\cite{anto3,afgnt}
to calculate the 
one loop correction to the prepotential of $N{=}2$ orbifold 
compactifications of the heterotic string. 
While the one loop prepotential has been calculated with the use 
of string amplitudes in \cite{afgnt}) via its third derivative with
respect to the T moduli, here we will provide 
an alternative 
way of calculating the one-loop correction\cite{wkll,afgnt}
to the prepotential
of the vector multiplets of the $N{=}2$ orbifold compactifications
of the heterotic string.
For the calculation of the one-loop contribution to the K\"{a}hler metric
we use the linear multiplet formulation\cite{cefergi1,anto3} and 
not the chiral formulation.
Note that both formulations are equivalent, since the linear multiplet
can always be transformed in to a chiral multiplet by a 
supersymmetric duality transformation.

In the superconformal formalism\cite{kuue}, the action for the 
linear multiplet is given up to one loop order by
\begin{equation}
{\cal L}=-{(S_o {\bar S}_o)}^{3/2}(\frac{\hat{L}}{2})^{-\frac{1}{2}}
e^{-\frac{G^{(o)}}{2}} + (\frac{\hat{L}}{2})G^{(1)} +{(S_o^3 w)_F},
\label{linear1}
\end{equation}
where now the gauge kinetic function is given by
$G^{(o)}(z,{\bar z}) + lG^{(1)}(z,{\bar z})$.
The vev of $l$ is the four 
dimensional gauge coupling constant $g^2$. 
Here,  $\hat{L}$ is the real linear multiplet 
satisfying ${\cal D}^2 \hat{L} =0$.

Eqn.(\ref{linear1}) does not have the the gravitational kinetic energy
$\propto R$ term to its canonical form. Instead, the chiral compensator
field is used to properly normalising its coefficient, procedure which
fixes the value of the compensator field.
The advantage of using the linear multiplet instead of the chiral 
multiplet is that it provides
an easy way of calculating\cite{anto3} one loop corrections 
to the K\"{a}hler metric. An easy way to see this comes from the 
following equation\footnote{Coming by expanding 
eqn.(\ref{linear1}).}, which includes the
bosonic kinetic energy terms,
\begin{equation}
{\cal L}_{bosonic}=-\frac{1}{4 l^2}\partial_{\mu}l \partial^{\mu}l
+\frac{1}{4 l^2}h^{\mu}h_{\mu}-G_{i{\bar j}} \partial_{\mu} z^i
\partial^{\mu} {\bar z}^{\bar j} -\frac{i}{2}(G_{ij}\partial z^j-
G_{i{\bar j}} \partial_{\mu}z^j -G_{i{\bar j}}\partial_{\mu}
 {\bar z}^{\bar j})h^{\mu}.
\label{linear2}
\end{equation}
The last term in eqn.(\ref{linear2}) reveals that the one loop 
correction to the K\"{a}hler metric $G_{z,{\bar z}}$ will come
by calculating the CP-odd part of the amplitude between the complex 
scalars and the antisymmetric tensor $b^{\mu \nu}$
\begin{equation}
<z(p_1){\bar z}(p_2)b^{\mu \nu}(p_3)>_{odd} = i \epsilon^{\mu \nu
\lambda \rho}p_{1\lambda}p_{2\rho}G^{(1)}_{z{\bar z}}.
\label{correk1}
\end{equation}
Here, G is the
K\"{a}hler metric and
$h^{\mu}=\frac{1}{2} {\epsilon}^{\mu \nu \lambda \rho}
\partial_{\nu} b_{\lambda\rho}$
is the dual field strength of the antisymmetric tensor field
$b_{{\lambda}{ \rho}}$.

The amplitude receives contributions only from $N{=}2$ sectors.
We are not considering contributions to the 
K\"{a}hler metric 
which arise from $N=1$ sectors, since these contributions  
arise only in $N{=}1$ orbifold compactifications of the  
heterotic string.  Here, we are only interested in the geometry
underlying the $N=2$ sectors.
Remember, than in general string amplitudes involve not only diagrams
that can be characterized 1PI diagrams in comparison with the 
corresponding effective field theory diagrams. Rather, they include  
one-particle reducible ones. 
Integrating out heavy fields in effective field theories leaves
terms associated with non-renormalizable interactions. In string 
theory, integration of massive states produces holomorphic quantities
in the effective string action. The presence of additional massive
fields becoming massless in a region of moduli space, produces
non-holomorphic quantities, e.g a logarithmic singularity at weak
Wilsonian gauge coupling\cite{kalou1}.

Lets us suppose that the internal six dimensional lattice 
decomposes into $T^2 \oplus T^4$.
 The $T^4$ part of the lattice, may represent
an orbifold limit\cite{hemo,walt} of $K_3$ while the $T^2$ part,
which contains the usual T, U moduli may contain\footnote{As it happens
in $N=2$ $(4,0)$ models\cite{afgnt,walt}.}
 a lattice shift,
necessary for modular invariance.
In this subspace of the Narain moduli space, we want to calculate the 
moduli dependence
of the one loop correction to the prepotential. 
Denote the untwisted moduli from a $N=2$ sector 
by P, where P can be the T or U moduli
parametrizing\cite{dis} the two dimensional unrotated plane.
Then the one loop contribution\cite{anto3} to the 
 K\"{a}hler metric is given by 
\begin{eqnarray}
G^{(1)}_{P\;{\bar P}}=-\frac{1}{(P - {\bar P})^2} {\cal I},\;\;\;
{\cal I}= \int_{\cal F} \frac{d^2\tau}{{\tau}^2_2} \partial_{\bar \tau}
(\tau_2 Z) {\bar F}({\bar \tau}).
\label{metrik1}
\end{eqnarray}
Here, the integral is over the fundamental domain,
and the factor $-\frac{1}{(P - {\bar P})^2}$ is the tree level moduli 
metric $G^{(0)}_{P\;{\bar P}}$. Z is the partition
function of the fixed torus
\begin{equation}
Z = \sum_{(P_L,P_R) \in \Gamma_{2,2}}
q^{P_L^2 /2} {\bar q}^{P^2_R /2} ,\;q \equiv e^{2 \pi \tau} \;
\tau = \tau_1 + \tau_2 ,
\label{partik1}                                                                
\end{equation}
and $P_L, P_R$
are the left and right moving momenta associated with this plane.
$F(\tau)$ is a moduli independent meromorphic form\footnote{A 
function f is meromorphic at a point A if 
the function h, $ h(z) \stackrel{def}{=} (z- A) f(z)$ is 
holomorphic (differentiable) at 
the point A. In general, this means that the function h is 
allowed to have poles.}   
of weight $-2$ with a single pole at infinity due to the tachyon
at the bosonic sector.
The function F was fixed in \cite{afgnt} to be 
\begin{equation}
F(\tau)= -(1/\pi)\frac{j(\tau) [j(\tau) -j(i)]}{j_{\tau}(\tau)},\;\;\;
\;\;\;
j_{\tau }\stackrel{def}{=}\frac{\partial j(\tau)}{\partial \tau},
\label{partik2}
\end{equation} 
where j the modular function for the group $SL(2,Z)$.   
The function $F(\tau)$ is actually the index in the Ramond sector in the
in the remaining superconformal blocks. With the use of the relations
between $E_4$, $E_6$, $\triangle$ and j, appearing in the Appendix,
 we can easily see that F becomes
\begin{equation}
F(\tau)= -2  i \frac{1}{(2 \pi)^2} \frac{E_4(\tau) E_6(\tau)}
{\triangle(\tau)}.
\label{partiku2341}
\end{equation}

{\em  Prepotential of vector multiplets/K\"ahler metric }
 
The convention for the complex dilaton is $<S>= \frac{\theta}{\pi} + i
\frac{8 \pi}{g_s^2}$, where $g_s$ is the four dimensional string
coupling and the $\theta$ angle.
For the calculation of the prepotential of the vector multiplets
we will will follow the approach of \cite{afgnt}.
Recalling the general form of the K\"ahler potential
\begin{equation}
K= -\ln (iY),\;Y= 2(F- {\bar F}) - (T -{\bar T})((F_T + 
{\bar F}_{\bar T}) -(S - {\bar S})(F_S + {\bar F}_{\bar S}),
\label{partiko224}
\end{equation}
\begin{equation}
F = STU + f(T,U)+ f^{non-pert}.
\label{partik3}
\end{equation}
In the following we will neglect the non-perturbative contributions
$f^{NP}$ to f.
The lagrangian (\ref{linear1}) may be related to the chiral 
multiplet one , by a duality transformation.
We introduce the dilaton S as a Lagrange multiplier into
(\ref{linear1}), e.g $({\cal L} -L i(S-{\bar S})/4)_D$. 
Using the equation of motion for S we get
\begin{equation}
({\cal L} -L \partial {\cal L})_D \equiv \frac{-3}{2}S_o {\bar S}_o
e^{-\frac{K}{3}}.
\label{partik21}
\end{equation}
In this form the K\"{a}hler potential has an expansion
as
\begin{equation}
K=-\ln\{i(S - {\bar S}) -2iG^{(1)} \} + G^{(o)}
\label{partik31}
\end{equation}
The quantity $G^{(1)}$ is identified with the Green-Schwarz term 
at the semiclassical weak coupling limit $S \rightarrow \infty$.
Expanding ({\ref{partik31}) 
\begin{eqnarray}
K^{(1)}_{P{\bar P}}=\frac{2i}{(S-{\bar S})}G^{(1)}_{P{\bar P}},\;\;
\;\;G^{(1)}_{T\;{\bar T}}=\frac{i}{2(T-{\bar T})^2}
\left(\partial_T -\frac{2}{T-{\bar T}}\right)\left(\partial_U - 
\frac{2}{U - {\bar U}}\right)f + c.c.
\label{partik4}
\end{eqnarray}
Using the equations for the momenta
\begin{equation}
{p_L}=\frac{1}{\sqrt{2ImT ImU}}(m_1+ m_2{\bar U} + n_1 {\bar T}+n_2 
{\bar U}
{\bar T}),\;p_R =\frac{1}{\sqrt{2ImT ImU}}(m_1+ m_2{\bar U}+ n_1 T+
n_2 T{\bar U}),
\label{partik5}
\end{equation}
we can prove that $ \cal I$ satisfies\cite{anto3} the following
differential equation
\begin{equation}
\{\partial_T \partial_{\bar T}+\frac{2}{(T-{\bar T})^2}\}{\cal I}=
-\frac{4}{(T-{\bar T})^2}\int d^2\tau {\bar F}({\bar \tau})
\partial_{\tau} (\partial_{\bar \tau}^2+\frac{i}{\tau_2} 
\partial_{\bar \tau})
(\tau_2 \sum_{P_L,P_R} q^{P_L^2 /2} {\bar q}^{P^2_R /2}).
\label{partik6}
\end{equation}
The integral representation of eqn.(\ref{partik6}) is a total 
derivative 
with respect to $\tau$ and thus zero.
However, the integral can give non-vanishing contributions at the 
enhanced symmetry points $T{=}U$. 

The general solution of (\ref{partik6}) away of the enhanced symmetry
points is\cite{afgnt}
\begin{equation}
{\cal I}=\frac{1}{2i} (\partial_T - \frac{2}{T-{\bar T}})
(\partial_U  - \frac{2}{U - {\bar U}} )f(T,U) + c.c
\label{partik7}
\end{equation}
Note that 
f represents the one-loop correction to the prepotential of the 
vector multiplets T, U and determines via eqn.(\ref{partik4}) the one
loop correction to the K\"{a}hler metric for the T, U moduli.
While $f_{TTT}$ was calculated for 
$N=2$ compactifications of the heterotic string in $D=4$
through its modular properties\cite{wkll} and from string 
amplitudes\cite{afgnt}, f could only be calculated indirectly\cite{mouha}.
Up to now the only way of calculating f for heterotic string
compactifications was indirectly\cite{mouha}, through the
one loop corrections to the Wilsonian gauge couplings\cite{sv,sv1,kj}} 
in the form
\begin{equation}
\partial_U \partial_{\bar U} \triangle = bK_{U{\bar U}}^{(o)} +
4 \pi^2 K_{U {\bar U}}^{(1)}.
\label{partik777}
\end{equation}
Here, $\triangle$ is given by\cite{anpatay,kap,anto1,dkl2}
\begin{equation}
\triangle =-\frac{1}{2} \int_{\cal F}\frac{d^2 \tau}{\tau_2} \frac{1}{\eta^2
({\bar \tau})} Tr_R F (-)^F (T_a^2 -\frac{1}{2 \pi \tau_2}),
\label{partik77771}
\end{equation}
where the trace is over the R-R sector, $T_a$ is the gauge group generator
of the gauge group factor $G_a$, F is the worldsheet fermion number. 
In addition, $K_{U{\bar U}}^{(o)}$ is the tree level K\"ahler 
metric\cite{anto3,mouha}
$-1/(U-{\bar U})^2$ and $K_{U {\bar U}}^{(1)}$ is the one loop K\"ahler metric 
given by 
\begin{eqnarray}
K_{U {\bar U}}^{(1)}= -\frac{1}{8 \pi^2} \int_{\cal F} \frac{d^2 \tau}{\tau_2}
\partial_U \partial_{\bar U} \left(
  \sum_{
 BPS\;hypermultiplets} - \sum_{ 
 BPS\;vector\;multiplets} \right) e^{i \pi \tau M_L^2} e^{-i \pi {\bar \tau}
  M_R^2},
 \label{kirios1}
 \end{eqnarray}
where $M_L$, $M_R$ are the masses for the left and right movers
and $\bigtriangleup$ is defined in terms of gauge couplings as
\begin{equation}
\frac{4 \pi^2}{g^2} = \frac{\pi}{2}Im S + \triangle.
\label{kiriooos11}
\end{equation}

In \cite{afgnt} it was shown function $f(T,U)$ of (\ref{partik7})
satisfies the differential equation
\begin{equation}
-i(U -{\bar U}) D_T \partial_T \partial_{\bar U} 
{\cal I}=\partial_T^3 f,
\label{partik8}
\end{equation}
where $D_T = \partial_T + \frac{2}{(T - {\bar T})}$ is the 
covariant derivative. 
Expansion of the l.h.s and integration by parts results in
\begin{equation}
f_{TTT}=4 \pi^2 \frac{U-{\bar U}}{(T -{\bar T})^2}\int d^2 \tau
{\bar F}({\bar \tau})\sum_{P_L,P_R} P_L {\bar P}_R^3
 q^{P_L^2 /2} {\bar q}^{P^2_R /2}.
\label{partik9}
\end{equation}
Examination of the behaviour of the r.h.s of eqn.(\ref{partik9}) under 
separately modular transformations $SL(2,Z)_T$ , $SL(2,Z)_U$, together
with examination of its singularity structure at the enhanced symmetry
point $T{=}U$, uniquely determines the well known solution
of the third derivative of the vector multiplet prepotential.
Remember that we examine the behaviour of the prepotential
including the region of the moduli space where we have gauge symmetry
enhancement to $U(1) \times SU(2)$. 

For
$N{=}2$ heterotic strings compactified on decomposable orbifolds
 $f_{TTT}$ was found to be \cite{afgnt} 
\begin{equation} 
f_{TTT}= -{2i\over\pi}\frac{j_T(T)}{j(T)-j(U)}
\left\{\frac{j(U)}{j(T)}\right\}
\left\{\frac{j_T(T)}{j_U(U)}\right\}
\left\{\frac{j(U)-j(i)}{j(T)-j(i)}\right\}.
\label{partik10}
\end{equation}
In \cite{wkll,afgnt} $f_{TTT}$ was determined by the property
of behaving as a meromorphic modular form of weight 4 under 
T-duality. 
In addition, $f_{TTT}$ had to vanish at the order 2 fixed point
$U{=}i$ and the order 3 fixed point $U{=}\rho$ of the 
modular group $SL(2,Z)$. Moreover, it had to transform with modular 
weight $-2$ under $SL(2,Z)_U$ transformations and
exhibit a singularity at the $T{=}U$ line.

Here, we will complete this picture by giving more details
of the calculation.    
Lets us denote the function with modular weight 4 under T-duality 
 $F^{(4)}(T)$ and the function with modular weight -2 under U-duality and
the singularity at $T{=}U$ by $F^{(-2)}(U)$. Then we must have 
\begin{equation}
F^{(4)}(T) = \frac{j_T^2}{j(j-1)},\;\;\;\;and\;\;\; 
F^{(-2)}(U)=\frac{j(U) (j(U) - j(i))}{j_U(U)(j(U) - j(T))}.
\label{partik101}
\end{equation}
As we can regognize $F^{(4)}(T)$ is the $E_4(T)$
function which is part of the basis of the modular forms
for the group $SL(2,Z)$. 
Moreover, $F^{(-2)}(U)$ can be rewritten in turn as 
\begin{equation}
F^{(-2)}=\frac{1}{j(T)-j(U)}\{- \frac{j_U^2(U)}{j(U)(j(U)-j(i))}\}
\frac{j_U^3(U)}{j(U)^2(j(U)-j(i))}(\frac{j_U^6}
{j^4(j(U)-j(i))^3 })^{-1}.
\label{partik102}
\end{equation}  
The terms appearing in eqn.(\ref{partik102}) represent
\begin{equation}
F^{(-2)}(U)= const.\;
\{\frac{1}{j(T)-j(U)}\}\;E_4(U)\;E_6(U) \eta^{-24}(U),
\label{partik103}
\end{equation}
where the standard theorems of modular forms predict
\begin{eqnarray}
E_4(U) \propto \frac{-j_U^2(U)}{j(j(U)-j(i))},\;E_6(U) \propto 
\frac{J_U^3(U)}{
j^2(U)(j(U)-j(i))},\;
\eta^{24} \propto \frac{j_U^6(U)}{j^4(U)(j(U)-j(i))}.
\label{partik104}
\end{eqnarray}
The function $E_4$ has a zero at $T{=}\rho$, and $E_6$ has a zero 
at $T{=}1$.
In addition, $\eta(T)$ is the well known cusp form the Dedekind
function. It has a zero at $T{=}\infty$.
Examination of the integral representation of behaviour of $f_{TTT}$ 
near the point $T{=}U$, shows that it has a single pole with 
residue $\frac{-2i}{\pi}$. This therefore 
fixes the   
numerical coefficient in front of $F^{(-2)}(T) \times F^{(4)}(U)$. 
Together with $F^{-2}(U)$, $F^{4}(T)$, 
we get the correct result eqn. (\ref{partik10}). 

{ \em The prepotential $f_{UUU}$ }

The prepotential function for the $f_{UUU}$ can be obtained by the 
replacement $T \leftrightarrow U$ but we prefer to 
find it from the equation for $f_{UUU}$. By taking appropriate
derivatives on (\ref{partik7})
we find
\begin{equation}
\partial_{U}^3f = f_{UUU}= -i(T-{\bar T})^2 \partial_{\bar T}D_U 
\partial_U {\cal I},
\label{oux1}
\end{equation}
or equivalently
\begin{equation}
\partial_{U}^3 f = i (T- {\bar T})^2 \partial_{T} 
D_U \partial_U {\cal I}.
\label{ouxc1}
\end{equation}
Here $D_U=\partial_U +\frac{2}{U-{\bar U}}$, the covariant derivative with
respect to U variable. 
It transforms with modular weight 2 under $SL(2,Z)_U$ modular transformations,
namely
\begin{eqnarray}
U\stackrel{SL(2,Z)_U}{ \rightarrow} \frac{ aU + b}{cU + d},\;\;\;\;D_U  
\rightarrow (cT+d)^2 D_U.
\label{covartibatt}
\end{eqnarray}

We should notice here, that because of the symmetry exchange $T
\rightarrow U$, the result for $f_{UUU}$ may come directly from
(\ref{partik10}), by the replacement $T \rightarrow U$. However, this 
can be confirmed by the solution of (\ref{oux1}).  
By using the explicit form of the expression (\ref{metrik1}) 
and the values
of the lattice momenta (\ref{partik5}), we evaluate the 
right hand side of (\ref{oux1}) as 
\begin{equation}
f_{UUU} = 8 \pi^2 \frac{(T-{\bar T})}{(U-{\bar U})^2} \int 
\frac{d^2 \tau}{\tau_2^2} {\bar F}({\bar \tau}) \partial_{\bar \tau}
\left( \tau_2^2 \partial_{\tau} (\tau_2^2 \sum_{P_L, P_R}P_L P_R P_R P_R
e^{\pi i \tau |P_L|^2} e^{-\pi i {\bar \tau} |P_R|^2} )\right).
\label{korikou1}
\end{equation}
Further integration by parts, with the boundary term vanishing away from 
the enhanced symmetry points, gives  
\begin{equation}
f_{UUU}= 4 \pi^2 \frac{(T-{\bar T})}{(U-{\bar U})^2} \int 
\frac{d^2 \tau}{\tau_2^2} {\bar F}({\bar \tau}) 
\sum_{P_L, P_R} P_L P_R P_R P_R
e^{\pi i \tau |P_L|^2} e^{-\pi i {\bar \tau} |P_R|^2}.
\label{korikou2}
\end{equation} 
Using now, the modular transformations of the momenta
\begin{eqnarray}
(P_L, {\bar P}_R)\stackrel{SL(2,Z)_T}{\rightarrow} \left({\frac{cT+d}{
c{\bar T}+d}}\right)^{\frac{1}{2}} (P_L, {\bar P}_R),\;\;
(P_L, {\bar P}_R)\stackrel{SL(2,Z)_U}{\rightarrow}
\left({\frac{cU+d}{
c{\bar U}+d}}\right)^{\frac{1}{2}} (P_L, {\bar P}_R),
\label{oux5}
\end{eqnarray}
we can see that $f_{UUU}$ transforms correctly, as it should, that is
modular weight 4 under $SL(2,Z)_U$ and $-2$ under $SL(2,Z)_T$. 
In addition, we can observe, in analogy with $f_{TTT}$,  
that (\ref{korikou2}) possesses a simple pole at $T=U$. 
The $f_{UUU}$ function has similar modular properties,
equivalent under the exchange monodromy transformation symmetry 
of f, and singularity structure as $f_{TTT}$.

As a result, $f_{UUU}$ takes the form
\begin{equation}
f_{UUU} = \frac{ (j(T) - j(i)) j(T)}{j_T(T) (j(T) - j(U))} \Psi(U),
\label{srty1}
\end{equation}
with $\Psi(U)$ a meromorphic modular form of weight 4 in U.
The T dependent part of $f_{UUU}$ is a meromorphic modular form of
weight $-2$ and has a singularity at the point $T=U$.

From the integral representation of $f_{UUU}$, eqn. (\ref{korikou1})
we can see that at the limit $T \rightarrow \infty$, $f_{UUU} 
\rightarrow 0$,
which means that $\Psi(U)$ is holomorphic anywhere.
Finally, we get
\begin{equation}
f_{UUU} = - \frac{2i}{\pi} \frac{ (j(T) - j(i)) j(T)}{j_T(T) (j(T) - j(U))}
\frac{j_U^2(U)}{j(U) - j(i)}\frac{1}{j(U)}.
\label{srty2}
\end{equation}
Lets us check the behaviour of (\ref{srty2}) away and at the fixed
points.
Away form the fixed points, e.g when $U \rightarrow {\hat T}= 
\frac{ a T + b}{c T + d}$, $f_{UUU}$ exhibits a singularity 
\begin{equation}
f_{UUU} \rightarrow - \frac{2i}{\pi} \frac{1}{U -{\hat T}}( cT + d)^2,
\label{srty3}
\end{equation}
such that the one loop K\"ahler metric $G_{U{\bar U}}^{(1)}$ behaves exactly
as expected\footnote{From the symmetry enhancement point of view for the
$SL(2,Z)$ modular group of the two torus.}, namely
\begin{equation}
G_{U {\bar U}}^{(1)} \rightarrow \frac{1}{\pi} \ln |U- {\hat T}|^2
G_{U{\bar U}}^{(0)}.
\label{srty4} 
\end{equation}
Exactly, when T is one of the fixed points of the modular group
$SL(2,Z)$, $f_{UUU}$ vanishes.
The presence of the logarithmic singulariry in the one loop corrections
to f gives rise to
the generation
of the discrete shifts in the theta angles due to monodromies around
semi-classical singularities in the quantum moduli space where
previously massive states become massless\cite{seiwi1,wkll,NS}.

{\em The one loop prepotential }f

The one-loop contribution to the K\"{a}hler metric can be
calculated through an equation different than eqn.(\ref{partik8}).
Expanding (\ref{partik7}) appropriately, we get 
\begin{equation}
2i{\cal I} =  \partial_T\partial_U f(T,U) -\frac{2}{T- {\bar T}}f(T,U) -
\frac{2}{U - {\bar U}}f(T,U) + \frac{4}{ (T-{\bar T}) (U -{\bar U})}f(T,U).
\label{oux11}
\end{equation}
Acting with the appropriate derivatives on the left hand 
of (\ref{oux11}) the following identity holds:
\begin{equation}
2f= i(T-{\bar T})^2 (U-{\bar U})^2 \partial_{\bar U}\partial_{\bar 
T}{\cal I},
\label{oux2}
\end{equation}
or in the symmetric form
\begin{equation}
2f= i(T-{\bar T})^2 (U-{\bar U})^2 \partial_{\bar T} \partial_{\bar U}
{\cal I}.
\label{oux222}
\end{equation}
Explicitly, 
\begin{equation}
2f=i(T-{\bar T})^2 (U-{\bar U})^2 \partial_{\bar U}\partial_{\bar
T} \int_{\cal F} \frac{d^2\tau}{{\tau}^2_2} \partial_{\bar \tau}
(\tau_2 Z) {\bar F}({\bar \tau}).
\label{ouxx2}
\end{equation}
Eqn. (\ref{oux2}) is the {\em master} equation for the prepotential.
It calculates the one loop
prepotential of any, rank four, four dimensional $N=2$ heterotic string
compactifications.
As we can observe, 
the one loop correction to the holomorphic prepotential comes by 
taking derivatives of ${\cal I}$ with respect to the conjugate 
moduli variables
from which the holomorphic prepotential does not have any dependence.
In this way, we 
{\em always produce} the differential equation for the f function 
from the string amplitude. In addition, the solution of this equation 
calculates the one loop correction to the K\"ahler metric.  

The integral representation of (\ref{oux2}), after using the
explicit form of momenta (\ref{partik5}), is
\begin{equation}
f = -4 \pi (T-{\bar T}) (U-{\bar U}) \int \frac{d^2 \tau}{\tau_2^2}
{\bar F}({\bar \tau}) \partial_{{\bar \tau}} [ \tau_2^2 
\partial_{ {\bar \tau}} (\tau_2 \sum_{P_L, P_R}{\bar P}_L {\bar P}_L 
e^{i\pi\tau |P_L|^2} e^{-i\pi{\bar \tau}| P_R|^2}
)],
\label{oux331}
\end{equation}
where we have used the identity
\begin{equation}
\partial_{\bar T} \partial_{\bar U}  
Z = -\frac{4 i \pi \tau_2}{(T -{\bar T})
(U - {\bar U})} \partial_{{\bar \tau}} ( \tau_2 \sum_{P_L, P_R}{\bar
P}_L {\bar P}_L Z)
\label{oux332}
\end{equation}
and the relations
\begin{equation}
\partial_{\bar T} {\bar P}_L^2= \frac{{\bar P}_L P_R}{T -{\bar
T}}= \partial_{\bar T} {\bar P}_R^2.
\label{sxesi1}
\end{equation}

We can easily see that the one loop prepotential has the correct modular
properties, it transforms with modular weight $4$ in T and $-2$ in U.
Eqn.(\ref{oux2}) is the differential equation that the one loop
prepotential
satisfies. The solution of this equation determines 
the one loop correction
to the K\"ahler metric and the K\"ahler potential for $N=2$ orbifold 
compactifications of the heterotic string. 

Compactifications of the heterotic string on $K_3 \times T_2$,
appears to have the same moduli dependence on T and U moduli, for 
particular classes of models\cite{afgnt,kava,klm1,afiq1}.
Formally, the same routine procedure, namely {\em taking the 
derivatives with 
respect to the conjugate T and U moduli on} ${\cal I}$, can be applied 
to any heterotic string amplitude between two moduli scalars and  
antisymmetric tensor, in order to isolate from the general solution 
(\ref{oux2}) the term $f(T,U)$. 
The solution for $f_{TTT}$ in 
eqn.(\ref{partik10}) was derived for $N=2$ compactification of the  
heterotic strings in \cite{afgnt} via the modular properties of the one 
loop 
prepotential coming from the study of its integral representation
(\ref{partik9}). Specific application for the model based on the 
orbifold
limit of $K_3$, namely $T^4/Z_2$, equivalent to the $SU(2)$
instanton embedding $(24,0)$, was given in \cite{mouha}. 
At the orbifold limit of $K_3$ compactification of the heterotic string 
the Narain lattice was decomposed into the form $\Gamma^{22,6}=
\Gamma^{2,2} \oplus \Gamma^{4,4} \oplus \Gamma^{16,0}$. 
It was modded by a $Z_2$ twist on the $T^4$ part together with a $Z_2$ 
shift $\delta$
on the $\Gamma^{(2,2)}$ lattice. For reasons of level matching 
$\delta^2$ was chosen to be $1/2$. The unbroken gauge group for this model
is $E_8 \times E_7 \times U(1)^2 \times U(1)^2$.
By an explicit string loop calculation via the one loop gauge couplings
in \cite{mouha}, from where the one loop prepotential was extracted with
an ansatz, they were able to calculate the third derivative of the
prepotential. The latter result was found to agree with the 
corresponding calculation in \cite{wkll,afgnt},
 which was calculated for the S-T-U subspace  
of the vector multiplets of the orbifold compactification of the 
heterotic string.
Here, we will check part of this result.
In particular, we will confirm the moduli dependence coming from
the trilogarithm, for the case of $SU(2)$ instanton 
embeddings\cite{lla1,fors}
$(12,12)$, $(10,14)$, $(11,13)$ and $(24,0)$ for which the 
index in the Ramond sector takes the same value. 

In reality, ${\bar F}({\bar \tau})$ is the trace of ${F}^{\prime} 
(-1)^{F^{\prime}} q^{L_o - 
\textstyle{\frac{c}{24}}} {\bar q}^{{\bar L}_o -
\textstyle{\frac{c}{24}}}/{\eta}({\bar \tau})^2$ over the Ramond 
sector boundary conditions of the remaining superconformal blocks.
For the S-T-U model with instanton embedding $(d_1,d_2)=(0,24)$ 
their supersymmetric index was calculated\footnote{Especially, for the S-T-U
models
 the contribution to the index from the
different
instanton embeddings $d_1, d_2$ in the two $E_8$ factors is 
independent\cite{dose1}
from the specific instanton embeddings.}
 in \cite{mouha}
in the form
\begin{eqnarray}
Index = \frac{1}{\eta^2} {Tr_R} F^{\prime}{(-1)}^{F^{\prime}} q^{L_o - \frac{c}{24} }
{\bar q}^{{\bar L}_o -\frac{c}{24}}\; =\; -2i \frac{{\bar E_4}\;
{\bar E_6}}{{\bar \Delta}},\;\;\;
\frac{{\bar E_4}{\bar  E_6}}{{\bar \Delta}}= \sum_{n \geq -1} c_1 (n) 
{\bar q}^n.
\label{partik233}
\end{eqnarray}
where $F^{\prime}$ is the right moving fermion number. 
This is exactly, the value of our index in eqn.(\ref{partiku2341}) except
for our normalization 
factor of $1/(2\pi)^2$ which accounts for the linear representation for
the dilaton.

Expanding $\cal I$ we get that
\begin{equation}
{\cal I}=(-i \pi)\int \frac{d^2 \tau}{\tau_2}(p_R^2 -\frac{1}{2\pi \tau_2})
{\bar F}({\bar \tau}).
\label{polik1}
\end{equation}
We remind here, a general remark, that the index ${\bar F}$ was  
determined using, 
the theory of modular forms, its modular properties and singularity 
structure alone.

Specific tests of dual pairs were performed, in the spirit of 
\cite{mouha}, in \cite{dose1,dose2,dose3,dose4,dose5}.
The low energy $N=1$ supergravity of type I and heterotic string 
theories
is subject to anomalies coming from hexagon diagrams
which prevent it from describing an anomaly free string theory. 
In this case anomalies are cancelled\cite{LWi,gs1,Calabi1} by the 
addition of
appropriate counterterms which modify the supersymmetry structure. 
Similarly, in six dimensions the total anomaly is associated to 
the eight form 
\begin{equation}
I_8 = {\tilde \theta}_1 tr R^4 + 
 {\tilde \theta}_2 (tr R^2 ) + {\tilde \theta}_3 tr R^2 tr F^2 
+ {\tilde \theta}_4  {{(tr F^2)}^2},
\label{eightf1}
\end{equation}
where
${\tilde \theta}_1$, ${\tilde \theta}_2$, ${\tilde \theta}_3$,
 ${\tilde \theta}_4$ are 
numbers depending on the spectrum\cite{gswe,erle} of the theory. 
Cancellation of the anomaly requires 
$\textstyle{{\tilde \theta}_1 = n_H - n_V + 29 T -273 = 0}$, where
$n_V$, $n_H$, $n_T$ are the numbers of vector multiplets, 
hypermultiplets and antiselfdual tensor multiplets respectively.
Because in six dimensions we have one tensor 
multiplet\cite{sost,cla,hiov,toira},
which incorporates the dilaton, a Weyl spinor and an antiself-dual
antisymmetric tensor, the last constraint becomes $n_H - n_V =244$.
Now Green Schwarz mechanism factorization of anomalies is at work
with $I_8 \propto -{\cal G}
{\tilde {\cal G}}$, ${\cal G} = tr R^2 - \sum_a \upsilon_a (tr F^2)$ 
and\footnote{Here, R, F are the gravitational and gauge field
strengths. 
The coefficients $\upsilon_a$, ${\tilde \upsilon}_a$ depend on the
particle content and the sum is over the gauge group G factors $G_a$.}
${\tilde {\cal G}} = tr (R \wedge R) - \textstyle{\sum_a 
{\tilde \upsilon}_a } tr (F \wedge F)_a$.
Cancellation of anomalies requires modification of the antisymmetric field 
stregth H as
\begin{equation}
H=dB + \omega^L - \sum_a {\upsilon}_a \omega_a^{YM},\;
\omega_L = tr (\omega R - \frac{1}{3} \omega^3),\;
\omega^{YM} = tr(A F - \frac{1}{3}\omega^3).
\label{modif1}
\end{equation} 
Here, $\omega_L$ , $\omega^{YM}$
are the Yang-Mills and Lorentz Chern-Simons three forms,
A the gauge field, R the Riemann tensor and $\omega$ the spin 
connection. However, because H is globally defined on $K_3$,
$\textstyle{\int_{K_3} dH=0}$.  As a result, 
we get that the following constraint has to be satisfied, 
\begin{equation}
\sum_a n_a = 24,\; n_a = \int_{K_3} tr F^2,\; \int_{K_3}tr R^2 =24.
\label{modify2}
\end{equation}
Here, the instanton number $n_a$ becomes equal to the Euler number of 
$K_3$.  
Intially, in ten dimensions the unbroken group is $E_8 \times  E_8 \times 
U(1)^4$,
where the $U(1)$'s are associated with the $T^2$ and the graviton and the 
graviphoton.  
The spectrum of the theory after compactification on $K_3 \times T^2$
can be calculated\cite{gswe} using index theory\cite{walt,kava}.
The gauge group G can be broken to a subgroup H, 
by vacuum expectation values of $K_3$ gauge fields in $\cal G$, where 
$H \times {\cal G} \subset G$.
The gauge group G breaks into the subroup H,  
which is the maximal subgroup commuting with the $\cal G$ subroup,
the commutant of $\cal G$.
We perform the decomposition adj$ G = \sum_{i} (R_i, M_i)$, where
$R_i$, $M_i$ representations of the gauge groups H and $\cal G$ 
respectively.
Then the number of left-handed spinor multiplets transforming in the
$R_i$ representation of H is given by 
\begin{equation}
N_{R_i}=\int_{K_3} -\frac{1}{2}tr_{R_i}F^2 + \frac{1}{48} dim_{M_i}
tr R^2 = dim_{M_i} -\frac{1}{2} \int_{K_3} c_2(V) index(M_i),
\label{index1}
\end{equation}
where V is the $\cal G$ bundle parametrizing the expectation
values(vev's) of the vacuum gauge fields on $K_3$. By $c_2(V)$ we 
denote the
second Chern class of the gauge bundle V and $dim_i$ the dimension
of the representation i.
In addition, the dimension of the moduli space of gauge bundles is
$4 h_a - dim({\cal G}_a)$, where $h_a$ is the Coxeter number of
${\cal G}_a$ and dim its rank. 
In a general situation we allow for the gauge group G to break to 
the commutant of $\otimes \cal G$, by embedding the gauge connections
of a number of a product of gauge bundles $V_a$ with gauge group 
${\cal G}_a$ into G, 
resulting
in the breaking of G into the commutant of $\otimes_a {\cal G}_a$.
In this way, we identify, for manifolds of $SU(2)$ holonomy, 
the spin connection of $K_3$ with the gauge group $\otimes_a 
{\cal G}_a$, breaking the G symmetry into H. This is the analog of
breaking the gauge group $E_8$, in manifolds of $SU(3)$ holonomy,
by the standard embedding\cite{iss1} of the $SU(3)$ gauge connection 
into the spin connection, to the 
phenomenologically interesting $E_6$ gauge group.

{\em Comments on the modular integral calculation}

Let us apply eqn.(\ref{ouxx2}) for the calculation of prepotential in the 
S-T-U model. 
Remember that the prepotential for this model was calculated from an ansatz 
solution. 
The index for this model is independent\cite{dose1} of the z
particular instanton embedding $(n_1, n_2)$ in the two $E_8$ factors
and is equal to (\ref{partik233}). We set 
\begin{equation}
\frac{E_4 E_6}{\triangle}({\bar \tau})= \sum_{n \geq -1}c(n)q^n=
c(-1)q^{-1} + c(0) + \dots
\label{tope1}
\end{equation}
The $\cal I$ integral in eqn.(\ref{oux2}) has been discussed before in
\cite{fors}. 
Using the values of the momenta (\ref{partik5})
in (\ref{oux2}) and using Poisson resummation we get
\begin{equation}
{\cal I}= (i \pi)T_2^2 \int 
\frac{d^2 \tau}{\tau_2^4} \sum_{n_1, n_2, l_1, l_2}
Q_R {\bar Q}_R e^{-2 \pi i {\bar T} det A}
e^{\frac{- \pi T_2}{\tau_2 U_2}|n_1 \tau + n_2 U \tau -U l_1 +l_2|^2}
{\bar F}({\bar \tau})
\label{tope2}
\end{equation}
where 
\begin{equation}
Q_R = \frac{1}{\sqrt{2 T_2 U_2}}(n_2 {\bar U}\tau+ n_1 \tau - {\bar U} l_1 +
l_2),\;\;\;
{\bar Q}_R = \frac{1}{\sqrt{2 T_2 U_2}}(n_2 U \tau + n_1 \tau - U  l_1 +
l_2).
\label{tope3}
\end{equation} 
The integral (\ref{tope2}) can be calculated using the method of
decomposition into modular orbits\cite{dkl2,kikou} of $PSL(2,Z)$.
There are three contributions to the modular integral.
The zero orbit $A=0$, the degenerate orbit and the non-degenerate
orbit.
The zero orbit $A =0$ gives no contribution to the $\cal I$ integral.
The next orbit that we will examine is the
non-degenerate orbit for which the matrix representative
is
\begin{equation}
A= \pm \left( \begin{array}{cc}
k&j\\
0&p
\end{array}\right),\;\; 0 \leq j < k,\;\;p \neq 0
\label{tope4}
\end{equation}
This integral has been calculated in \cite{fors} and it is
given\footnote{After proper
incorporation of the normalization factors of our Ramongd index.}
by
\begin{equation}
{\cal I}= -\frac{1}{(2 \pi)} \sum_{ \begin{array}{c}
k >0\\
l\in Z
\end{array}} \sum_{p>0} \delta_{n,kl}( \frac{2 kl}{p} +
\frac{l}{ \pi T_2 p^2} + \frac{k}{ \pi U_2 p^2} + \frac{1}{2 \pi^2 
 T_2 U_2 p^3} ) x^p + h.c,
\label{tope5}
\end{equation}
where $x \stackrel{def}{=} e^{2 \pi i (kT + lU)}$ and 
${\bar x} \stackrel{def}{=} e^{-2 \pi i(k {\bar T} + l {\bar U})}$.
Substituting eqn.(\ref{tope5}) in the {\em master equation }for the 
prepotential (\ref{oux2}) we get that the contribution of the 
orbit ${\cal I}_1$ in  f is
\begin{equation}
f|_{non-degenerate} = (2i) \left( \frac{2}{(2 \pi)^3} \sum_{(k,l)>0} c(kl)
{\cal L}i_3[e^{2 \pi i ( kT + lU)} ]\right).
\label{tope6}
\end{equation}
This is exactly the moduli dependence on the trilogarithm found 
indirectly in \cite{mouha}.
The dependence of the solution in i, out of the parenthesis in
(\ref{tope6}) is
neccessary since it is used to cancel the overall dependence on
i in the one loop K\"ahler metric (\ref{partik4}).  
Note that in the previous equation we have not considered the
complex conjugate solutions which arise by taking the partial
derivatives with respect to the $\bar T$ and $\bar U$ variables in the
complex conjugate part of the solution of eqn.(\ref{tope5}).
There are two ways to see this. One is the mathematical point 
of view while the other clearly come from physical requirements.
The physical point is that the prepotential has to be a holomorphic 
function of the vector moduli variables.
On the other hand,
the integral $\cal I$, which comes
as a solution of the one loop K\"ahler metric in eqn.(\ref{partik4}),
includes the complex conjugate part of the action of the
two covariant derivatives on the prepotential f.
However, the solution for the prepotential 
as was defined here in eqn.(\ref{oux2}) 
comes from the general
solution of the K\"ahler metric which does not include the 
conjugate part of its solution.
Results of the integration coming from the degenerate orbit
and related matters will appeal elsewhere\cite{kokos}.

\section{Application to rank three $N=2$ heterotic string compactifications}

We have said that one important aspect of the expected duality is that
the vector moduli space of the heterotic string must coincide at the 
non-perturbative level with the tree level exact vector moduli space 
of the type IIA theory. 
For the type IIA superstring compactified on a Calabi-Yau space
X the internal $(2,2)$ moduli space has $N=2$ world-sheet supersymmetry for
the left and the right movers and
is described, at the large complex structure limit of X, by
the K\"ahler\footnote{Let us consider the target space of a complex 
manifold
${\cal M}$ with dimension n. Choose coordinates on ${\cal M}$, $\phi_m$
and ${\bar \phi}_m$. Then ${\cal M}$ 
admits a
a K\"ahler structure if we can define a $(1,1)$ form J with the 
property $J= iG_{l {\bar m}} d\phi_m  \wedge d{\bar \phi}_l$ where for a 
 K\"ahler manifold the metric is $G_{l {\bar m}}= \partial_{\phi^m}
\partial_{{\bar \phi}_l}K$, and the  K\"ahler potential is K.}
moduli,  namely  $B+iJ \in H^2(X,C)$,
where $B+iJ = \sum_{i=1}^{h(1,1)} (B+iJ)_a e_a$ with $B_a$, $J_a$ real
numbers and $t_a =(B+iJ)_a$ representing the special coordinates and 
$e_a$ a basis of $H^2(X,C)$.

In this section we will derive the general form of the 
equation determining the prepotential form the rank three $N=2$ 
heterotic compactifications.
In particular we will examine a type II model admitting 
a heterotic perturbative dual realization. 
The heterotic model contains three moduli the dilaton S,  the graviphoton,
and one moduli
the T moduli. It coincides with the 
corresponding Calabi-Yau dual model at its weak coupling limit.

In order for the heterotic prepotential to match its Calabi-Yau dual 
at its weak coupling limit a number of conditions are necessary, which
we will briefly review them here.
The existence of a type II dual to the weak coupling phase 
of the heterotic string is exactly the existence of the 
conditions\cite{lla}
\begin{equation}
D_{sss}=0,\;\;\;\;D_{ssi} =0\; for\;every\;i,\;,   
\label{koufame12}
\end{equation}
where D the Calabi-Yau divisors appearing in (\ref{koufme1}).
An additional condition originates 
from the higher derivative gravitational couplings
of the heterotic vector multiplets and the Weyl multiplet
of conformal $N=2$ supergravity\cite{gavos1}. The relevant couplings
originate from terms in the form $g_n^{-2} R^2 G^{2n-2}$,  where
R is the Riemann tensor, G the field strength of the graviphoton.
The $g_n$ couplings obey $g_n^{-2}=Re {\tilde F}_n(S,M^i) + \dots$.
The same of couplings appear in type II superstring\cite{berda}. 
In the heterotic side they take the form
\begin{equation}
{\tilde F}_n = {\tilde F}^{(0)}(S, M_i) + {\tilde F}^{1}(M^i)
+ {\tilde F}^{NP}(e^{-8\pi^2 S}, M^i),\;\;{\tilde F}_1=24 S,\;\;
{\tilde F}^o_{n \geq 1} = const,
\label{koufame141}
\end{equation}
where S is the heterotic dilaton and $M^i$ the rest of the vector
multiplets moduli.
Such terms appear as well in the effective action of type II vacua 
and they have to match with heterotic one's due to duality.
In the large radius limit the higher derivative
couplings satisfy(the lowest order coupling) 
${\tilde F}_1 \rightarrow -\frac{4 \pi i}{12}\sum_{a} 
(D_a \cdot c_2) t_a $, 
where $c_2$ is the second Chern class of the three fold X.
Because at the tree level, $g_1^{2}= Re {\tilde F}_1$ we can infer the 
result that
\begin{equation}
D_a \cdot c_2(X) =24.  
\label{koufame15}
\end{equation}
The last condition represents\cite{lla} the mathematical
fact that the Calabi-Yau threefold X admits a fibration $\Phi$ such as
there is a map $X \rightarrow W$, with the base being $P^1$ and generic 
fiber the $K_3$ surface.
Furthermore, the area of the base of the fibration gives
the heterotic four dimensional dilaton.

In the content of the moduli of the Calabi-Yau space of X, the
holomorphic
prepotential at the large radius limit takes the form
\begin{equation}  
F= -\frac{i}{6}\sum_{\alpha,\beta,\gamma}(D_{\alpha} \cdot D_{\alpha}
\cdot D_{\gamma})t_{\alpha} t_{\beta} t_{\gamma}
- \frac{\chi \zeta(3)}
{2(2\pi)^3}\sum_{(d_i)_{i=1,\dots,n}} n_{d_1,\dots,d_n}
{\cal L}i_3 (\Pi_{i=1}^{n} q_i^{d_i}),
\label{koufme1}
\end{equation}
where the trilogarithmic function is 
${\cal L}i_3(x) \stackrel{def}{=} \textstyle{\sum_{j \geq 1}\frac{x^j}{j^3}}$. The
first term in eqn.(\ref{koufme1}) is a product of the the Calabi-Yau 
divisors D, associated to the basis $e_a$, 
and the second term\cite{can} represents world-sheet instanton 
contributions. The $n_{d_1,\dots,d_n}$ are the world sheet instanton 
numbers, the numbers 
of genus zero rational curves, and $d_i$ the degrees of the curves.
Performing duality tests, at the weak coupling heterotic limit, between a 
heterotic model and its possible
type IIA dual is then equivalent to comparing the weak coupling limit 
of the prepotential\cite{kava} of the vector multiplets for the 
heterotic string with the large radius limit of (\ref{koufme1}).
After identifying the heterotic dilaton with one of the vector moduli
of the type IIA model in the form $t_s =(B+iJ)_s = 4\pi i S$,
the type IIA prepotential takes the general form\cite{cande}
\begin{equation}
{\cal F}_{IIA} =-\frac{1}{6}C_{ABC} t^A t^B t^C -\frac{\chi \zeta(3)}
{2(2\pi)^3}\sum_{d_1\dots,d_n} n_{d_1,\dots,d_n} {\cal L}_{i_3}[e^{2\pi 
i[\sum_k d_k t^k]}],
\label{koufeme2}
\end{equation}
where we are working inside the K\"ahler cone\cite{can,hoso2} 
$\sigma\stackrel{def}{=}\{\sum_{\rho} t^{\rho} J_{\rho}|
t^{\rho}> 0 \}$, where $J^{\rho}$ generate the cohomology group 
$H^2(X,R)$ of the Calabi-Yau three fold X.

In \cite{klm1} it was noticed that the nature of type II-heterotic sting
duality test has to come from the $K_3$ fiber structure over $P^1$
of the type IIA side.                                                                                                                                    
The form of the $K_3$ fibration can be found\cite{klm1,afiq1} 
by taking for example
the CY in $P^4(1,1,2k_2,2k_3,2k_4)$ and then set $x_o =\lambda x_1$. 
After rescaling $x_1 \rightarrow x_1^{1/2}$ we arrive at the equation
for the hypersurface 
\begin{equation}
F(\lambda)Z_1^d +Z_2^{d/{k_2}} + \cdots = 0.
\label{koufeme4}
\end{equation}
The degree $d=1+k_2 + k_3 + k_4$. For generic values of $\lambda$ 
eq.(\ref{koufeme4}) is a $K_3$ surface in weighted $P^3$. So 
$P^4(1,1,2k_2,2k_3,2k_4)$ is a $K_3$ fibration
fibered over the $P^1$ base which is parametrized by $\lambda$.
At the large radius limit
of X  the heterotic dilaton S
is identified as one of the vector multiplet variables  
as $t_s= 4 \pi i S$.
Confirmation of duality between dual pairs is then  
equivalent to the identification\cite{louisne}
\begin{equation}
\textstyle{{\cal F}_{IIA}={\cal F}_{IIA}(t^s, t^i) +
{\cal F}_{IIA}(t^i)={\cal F}_{het}^o(S, \phi^I) +
{\cal F}_{het}^{(1)}(\phi^I)}.
\label{koufame11}
\end{equation}
Here,
we have expand the prepotential of the type IIA in its large radius
limit, namely large $t_s$. In the heterotic side, we have the 
tree level classical contribution 
as a function of the dilaton S and the vector multiplet moduli 
$\Phi^I$, in addition to the one loop correction as a function 
of only the $\Phi^I$.

Dual pairs for which the prepotential in the type IIA theory is known 
can be mapped to the type IIB using mirror symmetry\cite{grple}.
Let us review some aspects of the low energy theory of the type IIB
superstrings.
In Calabi-Yau manifolds, special geometry is associated with
the description of their moduli spaces. In type IIB,
the $H^{2,1}$ cohomology describes the deformation of the complex
structure of the Calabi-Yau space $\cal M$. Now the K\"ahler metric for
the $(2,1)$ moduli is defined\footnote{In the rest of the section
the notation for the special coordinates is as follows, $Z^i= -i
X^i/X^o$.}
 from the
Weyl Peterson metric\cite{can,cande,cande1,cgh}
$\sigma_{ij}$, namely
\begin{equation}
G_{ij} = \sigma_{ij}
/ (i(\int_{{\cal M}} \Omega \wedge {\bar \Omega})),
\label{homo1}
\end{equation}
where
\begin{equation}
{\varphi}_{i}=(1/2){\varphi}_{i k \lambda{\bar \rho}}dx^{k}dx^k
dx^{\lambda} dx^{\bar \rho},\;\;
\sigma_{ij} = \int_{{\cal M}} \varphi_i
\wedge {\bar \varphi}_{\bar j}
\label{homo2}
\end{equation}
and ${\varphi}_{ik\lambda{\bar
\rho}}= (\partial g_{{\bar \rho}{\bar \xi}}/ {\partial t^i})
{\Omega}^{\bar xi}_{k \lambda}$.
Here, $t_i=1,\dots,b_{2,1}$ and $G_{ij}=-\partial_i
\partial_j( i \int_{\cal M} \Omega \wedge {\bar \Omega})$.
The three form tensor $\Omega$ is given in terms of the
homology basis $\alpha$, $\beta$ as $\Omega = X^{I}\alpha_I + i F_I
\beta^{I}$.   
The complex structure is described by the periods of the holomorphic
three form $\Omega$ over the canonical homology basis.
Here, the periods are given by $X^I =\int_{A^I}\Omega$,
$iF_I =\int_{B^I}\Omega$ the integral of the holomorphic three form
over the homology basis. The K\"ahler potential comes from the
moduli metric
\begin{equation}
G_{ij}=-i\partial_i \partial_j \{ i \int \Omega \wedge {\bar
\Omega}\},\;\;
K=-\log(X^I {\bar F}_I + {\bar X}^I F_I).
\label{pol1}
\end{equation}
Now the Riemmann tensor is defined as
\begin{equation}
R_{i{\bar j}k{\bar l}}= G_{i{\bar j}} G_{k{\bar l}} +G_{i{\bar l}}G_{k{\bar
j}}-{\bar {\cal C}}_{ikn}{\bar {\cal C}}_{{\bar j}{\bar l}{\bar n}}
G^{n{\bar n}}e^{2K},  
\label{pol2}
\end{equation}
where the expression of the Yukawa couplings in a general coordinate
system are given by $ {\cal C}= \int \Omega \wedge 
\partial_i \partial_j   \partial_k \Omega $, $\partial_i = 
\partial / \partial z^i$.
The holomorphic function F does not receive quantum
corrections from world-sheet instantons and as a consequence neither the
the K\"ahler potential derived from it. 
Calabi-Yau threefolds can be constructed\cite{so} among other ways as 
a hypersurface or as a complete intersection of hypersurfaces in a
weighted
projective space $P^N(\vec w)$.
Remember, that the complex projective space $CP^N$ is the space defined
by the homogeneous complex coordinates $Z_1,\dots,Z_{N+1}$ which obey
$(Z_1,\dots,Z_{N+1}) \stackrel{\lambda \neq 0}{\equiv}(\lambda^{w_1}
Z_1,\dots,\lambda^{w_{n+1}} Z_{N+1})$ for complex $\lambda$.
The threefold is obtained from the $CP^4$, while the $K_3$ can be 
obtained from
the $\sum_{k_i} \alpha_{k_1 k_2 k_3 k_4} x_{k_1} x_{k_2} x_{k_3}
x_{k_4}=0$, in projective $P^3$ and $P^2$ respectively(rp).
They describe complex manifolds parametrized by 135 and 35
complex coefficients $a_{k_i}$ rp, which after removing an               
overall redundancy they give 101, 19 elements of $H^{(1,1)}$ rp. 
The weighted projective space  $P^N(\vec w)$ is defined by the conditions
on the 
homogeneous coordinates
$(Z_1,\dots,Z_{N+1})\stackrel{\lambda \neq
0}{\equiv}(\lambda^{d_1},\dots,\lambda^{d_{N+1}})$ and 
$P^N(\lambda^{d_1}Z_1,\dots,\lambda^{d_{N+1}}Z_{N+1})\stackrel{def}{=}
\{C^{N+1}/(Z_1=0,\dots,Z_{N+1}=0)\}$. The last condition, excludes 
the origin
of the complex space. The $d_i$ are the weights and the sum of the
weights is the degree of the variety.

Let us consider the Calabi-Yau three fold defined as the zero 
locus of the hypersurface $P^4_{1,1,2,2,2}$ of degree eight. 
This model appears in the list of \cite{kava} as 
the A model and it is defined as $X_8(1,1,2,2,2)^{-86}_2$, where the 
subscripts and
superscripts denote the Betti numbers $b_{1,1}=2$ and $b_{1,2}=86$.
This model gives rise to 2 vector multiplets
and 86 + 1 hypermultiplets including the dilaton and its moduli space
can be studied using mirror symmetry\cite{grple,hoson3}.

The mirror manifold $X_8^{*}$ for this model is defined
by the Calabi-Yau three fold in the form $\{{\cal P}=0\}/Z_4^3$,
where the zero locus is
\begin{equation}
{\cal P}=z_1^8 +z_2^8 +z_3^4 +z_4^4 + z_5^4 -8\psi z_1 z_2  
z_3 z_4 z_5 - 2\phi z_1^4 z_2^4.
\label{koili}
\end{equation}
It depends on the deformation parameters $\phi$ and $\psi$.
The $Z_4^3$ symmetry acts on the coordinates as $(z_2, z_{2+m})
\rightarrow (-iz_2, iz_{2+m})$ for $m=1,2,3,$ respectively.
A good description of the moduli space is obtained by enlarging 
the group $\{{\cal P}=0\}/Z_4^3$ to the group $\hat G$ acting as
\begin{equation}
(z_1, z_2, z_3, z_4, z_5 ;\psi, \phi)\rightarrow 
(\omega^{{\tilde a}_1}z_1, \omega^{{\tilde a}_2}z_2,\omega^{2{\tilde
a}_3}z_3, \omega^{2{\tilde a}_4}z_4, \omega^{2{\tilde a}_5}z_5;
\omega^{-{\tilde a}}\psi,\omega^{-4{\tilde a}}\phi),
\label{koili1}
\end{equation}
 where $\textstyle{{\tilde a}=e^{2\pi i/8}}$, ${\tilde a}_i$ are
integers such as ${\tilde a}={\tilde a}_1 +{\tilde a}_2 +
2 {\tilde a}_3+ {\tilde a}_4 + {\tilde a}_5 $.
Modding the weighted projective spaces by the group $\hat G$ requires
modding out by the action $(\phi, \psi) \rightarrow (-\phi,{\tilde
\alpha} \psi)$.
The prepotential of the type IIB model defined on the mirror manifold
$X_8^{*}$ was calculated in \cite{anpa} form the study of the Yukawa 
couplings in \cite{can,hoso1,hoso2} as
\begin{equation}
{\cal F}^{II}= - 2t_1^2 t_2 - \frac{4}{3}t_1^3 + \dots + f^{NP} .
\label{peroa}
\end{equation}
From the form of the prepotential we can infer that the type II
model has a heterotic dual which corresponds to the particular
identification of $t_2$ with the heterotic dilaton
and $t_1$ with the heterotic T moduli.
As a result
\begin{equation}
f^{heterotic} = -2 ST^2 + f(T) + f^{non-pertur},
\label{peroa1}
\end{equation}
where $f(T)$ the one loop correction and $f^{non-pertur}$ the 
non-perturbative contributions.
The heterotic model is an S-T model, a two moduli example or rank three 
model, if someone takes into account the graviphoton.
The  
exact correspondence of $P^4_{1,1,2,2,2}$  with the 
three rank heterotic model  
is their connection via their classical T-duality group.
The two models at the weak coupling limit of the $t_2$ moduli
have the same classical duality group, $\Gamma_o(2)_{+}$.
Study of the discriminant of $P^4_{1,1,2,2,2}$ gives that
the conifold singularity should correspond to the perturbative $SU(2)$ 
enhanced symmetry point.
This fixes the momenta for the $\Gamma^{(2,1)}$ compactification lattice
of the heterotic model as
\begin{equation}   
p_L = \frac{i \sqrt{2} }{T -{\bar T}} \left(n_1 +n_2{\bar T}^2 +
2 m {\bar T}\right),\;\;
p_R = \frac{i \sqrt{2}}{T -{\bar T}} \left( n_1 +n_2 T{\bar T} +
m (T + {\bar T})\right).
\label{koilhju}
\end{equation}
with the enhanced symetry point at level 2.
Let me discuss first the {\em master} equation for the general rank
three model, as well the dual heterotic of $P^4_{1,1,2,2,2}$.
The solution for the one loop correction to the K\"ahler metric 
reads\cite{anpa}
\begin{equation}
K_{T{\bar T}}=K_{T{\bar T}}^{(o)}\{1 + \frac{2i}{S- {\bar S}}{\cal I}+
\dots \}.
\label{atta}
\end{equation}
Using now the general form of solution for the K\"ahler metric
\begin{equation}
{\cal I} = \frac{i}{8} \left( \partial_T - \frac{2}{T -{\bar T}} \right)
\left( \partial_T - \frac{4}{T -{\bar T}}\right) f(T) + h.c,   
\label{koilikoili3}
\end{equation}
we can infer the {\em master} equation for the perturbative one loop
correction
to the prepotential as
\begin{equation}
2i  f(T) =  (T- {\bar T})^3 \partial_{\bar T} {\cal I}.
\label{koilikoili4}
\end{equation}
Here $K_{T{\bar T}}^{(o)}$ is the tree level metric $-2/(T-{\bar T})^2$
and ${\bar C}_l (\bar \tau)$ is the index of the
Ramond sector in the remaining superconformal blocks.
Note that eqn.(\ref{koilikoili4}) was derived from the general
solution for the one loop K\"ahler metric without any reference
to values of momenta for the $\Gamma^{(2,1)}$. This means that
this equation determines the prepotential for any rank three $N=2$
compactification of the heterotic string.
For example (\ref{koilikoili4}) determines the heterotic duals 
of the models B, C in \cite{klm1}, with associated modular groups
${\Gamma_o(3)}_{+}$, ${\Gamma_o(4)}_{+}$ 
and enhanced symmetry points at, 
the fixed points of their associated modular groups, 
Kac-Moody levels 3 and 4 respectively\cite{gome}.

The one loop K\"ahler metric\cite{anto3,anpa} for the 
heterotic model dual to the
type $P^4_{1,1,2,2,2}$ model reads
\begin{equation}
{\cal I}= \sum_{i=1}^{6}\int \frac{d^2 \tau}{\tau_2^{3/2}}
{\bar C}_l (\bar \tau) \partial_{\bar \tau}(\tau_2^{1/2}
\sum_{p_L,p_R \in \Gamma_l} e^{\pi i \tau |p_L|^2} e^{- \pi i
{\bar \tau}p_R^2}).
\label{koilikoili2}
\end{equation}
Direct substitution of the values  
of the $\Gamma^{(2,1)}$ compactification lattice momenta in (\ref{koilikoili2})
may give us the prepotential f in its integral representation.
Here the sum is over\cite{anpa} the different lattice sectors $\Gamma_l$,
\begin{equation}
\begin{array}{cccc}
\Gamma_{1,\epsilon}&\Gamma_{2,\epsilon}&\epsilon=1,\;2\nonumber\\
\Gamma_{1,\epsilon}&n_1 \in Z\ +\;\frac{1}{2}& n_2\;odd\nonumber\\
\Gamma_{2,\epsilon}&n_1 \in Z &n_2\;even\nonumber\\
\Gamma_{3,\epsilon}&n_1 \in Z + \frac{n_2 +1}{2}&n_2\;\in\; Z,
\label{adfgdg}
\end{array}
\end{equation}
where $m\;\in\;Z +\epsilon$, 
that are needed due to world-sheet modular invariance.

\section{Conclusions}

We have calculated the general equation which calculates directly
the one loop perturbative prepotential of $N=2$ heterotic string 
compactifications for any rank three or rank four parameter models.
These heterotic string compactifications may or may not have a 
type II dual compactified on a Calabi-Yau.
In general, heterotic vacua with instanton embeddings numbers
$(12-n, 12 + n)$ on the $E_8 \times E_8$ gauge bundle are 
associated\cite{ftheo,lla1} to the elliptic fibrations over the 
Hirzebrush surfaces $F_n$.          
Especially, for the families of Calabi-Yau threefolds with
Hodge numbers $(3,243)$, associated with $K_3$ fibrations and elliptic 
fibrations, when
n is even the rank three Calabi-Yau is an elliptic fibration over   
the Hirzebrush surface $F_2$ or $F_0$, while for n odd the rank three
models are given in terms of the Hirzebrush surface $F_1$.
At the heterotic perturbative level all the models, which are coming
from complete Higgising of the charged hypermultiplets,
with the same Hodge numbers, come from the instanton embeddings
$(12,12)$, $(11,13)$, $(10,14)$. 
However, at the heterotic perturbative level all the models are the 
same as
we have already said. This is clearly seen from the nature of the
Ramond index (\ref{partiku2341}) which is independent from the
particular instanton embedding.
In particular, we tested the moduli dependence of the
prepotential, coming from the non-degenerate orbit, for the
previous $SU(2)$ instanton embeddings, and the $(24,0)$ one, against the 
moduli dependence of the prepotential extracted from the one loop
corrections to the gauge couplings in \cite{mouha}.
In addition, we calculated the differential equation of the 
third derivative of the prepotential for the rank four S-T-U model
with respect of the complex structure U variable.  

The {\em master} equation's (\ref{oux2}), (\ref{koilikoili4})
 open the way for direct testing 
of the web of dualities, e.g 
the duality between type I and $K_3 \times T^2$ at their weakly coupled
region which was tested via the third derivatives of the one loop
prepotential in\cite{anbac}.
However, there are other dualities which can be tested at the quantum level. 
For example, if we continue\cite{vafos}, further compactification on $S^1$ of
F-theory
defined on the elliptic Calabi-Yau, we get duality between 
M-theory on the associated Calabi-Yau three fold and heterotic strings
on $S^1$. Further compactification, we get duality between type IIA on 
Calabi-Yau three folds and heterotic on $K_3 \times T^2$. 
Furthermore, the 
direct way of calculating the holomorphic prepotential in  
(\ref{oux2}), (\ref{koilikoili4}) 
can calculate the $N=2$
 central charge and $N=2$ BPS spectrum as well the black hole
entropy\cite{lubl1,reybl2}.

{\bf Acknowledgments}

We are grateful to I. Antoniadis, C. Kounnas, K. S. Narain, S.
Stieberger, D. Zagier and R. Lewes for 
useful conversations and D. Bailin for reading the manuscript.

{\bf Appendix A}\setcounter{equation}{0}
\def\theequation{A.\arabic{equation}} 

{\cal A1} {\em Useful relations with modular forms}

The functions $E_4$, $E_6$ form the basis of modular forms for
the group $SL(2,Z)$ and are defined in term of Eisenstein series 
of weight four and six. Namely,
\begin{equation}
E_{2k}(T)= \sum_{
n_1, n_2 \in Z}^{\prime}
(in_1 T + n_2)^{-2k},\;\;k\;\in\;Z.
\label{partik1041}
\end{equation}
Here the prime means that $n_1 \neq 0$ if $n_2 = 0$.
Let us provide some useful relations between the basis for
modular forms for $PSL(2,Z)$ and $\triangle$
and the j invariant.
With the use of these relations various results appeared in the 
literature, like those in \cite{wkll}, \cite{afgnt} can be easily
translated to each other.
It can be proved that the following relations hold
\begin{equation}
E_4(T) = -\frac{(j^{\prime})^2}{4 \pi^2 j(j-j(i))} = 1+240 
\sum_{n=1}^{\infty} \sigma_3(n) e^{2 \pi i T},
\label{areko1}
\end{equation}
\begin{equation}
E_6(T) =\frac{(j^{\prime})^3}{(2 \pi i)^3 j^2(j-j(i))}=
1-504\sigma_5(n) q^n,
\label{areko2}
\end{equation}
\begin{equation}
\triangle(T)= -\frac{1728(j^{\prime})^6}{(48 \pi^2)^3 j^4 (j-j(i))^3}= 
\eta^{24}(T)= \frac{1}{(2 \pi i)^6}\frac{(j^{\prime})^6}{j^4 (j-j(i))^3},
\label{areko3}
\end{equation}
where $\eta(T)$ is the Dekekind function
and the value of $\sigma$  represents the sum over divisors
\begin{equation}
\sigma_{h-1}(n)\stackrel{def}{=} \sum_{d/n} d^{h-1}.
\label{areko4}
\end{equation}
We have used the notation
\begin{equation}
j^{\prime}(T) = j_{T}(T).
\label{areko24}
\end{equation}
Note that in general $E_4$, $E_6$ and $\triangle$ are defined 
in terms of the Klein's absolute invariant J as
\begin{equation}
j(T)\stackrel{def}{=} 1728 J(T)
\label{areko44}  
\end{equation}
\begin{equation}
J(T)= \frac{E_4^3(T)}{1728
\triangle(T)}=
1 + \frac{E_6^2(T)}{1728 \triangle(T)},;\;\;\;T \in H
\label{areko5}  
\end{equation}
and 
\begin{equation}
j(T)=e^{-2 \pi i T} + 744 + 196884
e^{2 \pi i T} + \dots
\label{areko6}
\end{equation}
Remember that the following relations are valid
\begin{equation}
j(T)= \frac{E_4^3(T)}{\triangle(T)},
\label{areko7}
\end{equation}
and
\begin{equation}
j(T)= \frac{E_6^2(T)}{\triangle(T)}.
\label{areko8}
\end{equation}
Here, j is the modular invariant function for the
inhomogeneous  modular group
$PSL(2,Z)$.

\end{document}